\documentstyle[epsfig,mnras_cite,times]{mn}

\title{Weak lensing with COMBO-17: estimation and removal of intrinsic
alignments}
\author[Heymans et al.]{Catherine
Heymans$^{1}$\thanks{heymans@mpia.de}, Michael Brown$^{2}$,
Alan Heavens$^{2}$, Klaus Meisenheimer$^{3}$,\newauthor
Andy Taylor$^{2}$ \& Christian Wolf$^{1}$\\
$^1$ University of Oxford, Astrophysics, Keble Road, Oxford, OX1 3RH, UK \\
$^2$ Institute for Astronomy, University of Edinburgh, Royal Observatory,
Blackford Hill, Edinburgh, EH9 3HJ, UK \\
$^3$ Max-Planck-Institut f\"{u}r Astronomie, K\"{o}nigstuhl, D-69117,
Heidelberg, Germany}

\newcommand{\be}{\begin{equation}}  \newcommand{\ee}{\end{equation}}
  \newcommand{\ba}{\begin{eqnarray}}
\newcommand{\ea}{\end{eqnarray}}  
\newcommand{\nn}{\nonumber\\}

\newcommand{\bm}[1]{\mbox{\boldmath{$#1$}}}
\def\gs{\mathrel{\raise1.16pt\hbox{$>$}\kern-7.0pt %
\lower3.06pt\hbox{{$\scriptstyle \sim$}}}}         %
\def\ls{\mathrel{\raise1.16pt\hbox{$<$}\kern-7.0pt %
\lower3.06pt\hbox{{$\scriptstyle \sim$}}}}         %

\begin{document}

\maketitle

\begin{abstract}
We estimate and remove the contamination of weak gravitational 
lensing measurements by the intrinsic alignment of close pairs of 
galaxies.  We do this by investigating both the aperture mass B 
mode statistic, and the shear correlations of close and distant 
pairs of galaxies.  These can be used to quantify non-lensing 
effects in weak lensing surveys.  We re-analyse the COMBO-17 survey,
and study published results from 
the Red-sequence Cluster Survey and the VIRMOS-DESCART survey, 
concluding that the intrinsic alignment effect is at the lower 
end of the range of theoretical predictions.  We also revisit 
this theoretical issue, and show that misalignment of baryon and 
halo angular momenta may be an important effect which can reduce the 
intrinsic ellipticity correlations estimated from numerical 
simulations to the level that we and the SuperCOSMOS survey observe.  
We re-examine the 
cosmological parameter estimation from the COMBO-17 survey, using 
the shear correlation function, and now marginalising over the Hubble 
constant. Assuming no evolution in galaxy 
clustering, and marginalising over the intrinsic alignment 
signal, we find the mass clustering amplitude is 
reduced by 0.03 to $\sigma_8(\Omega_m / 0.27)^{0.6} = 
0.71 \pm  0.11$, where $\Omega_m$ is the matter density 
parameter.  We consider the forthcoming 
SuperNova/Acceleration Probe wide weak lensing survey (SNAP), and 
the Canada-France-Hawaii Telescope Legacy wide synoptic survey, and expect them
to be contaminated on scales $>1$ arcmin by intrinsic 
alignments at the level of $\sim 1\%$ and $\sim 2\%$ respectively.  
Division of the SNAP survey for lensing tomography significantly 
increases the contamination in the lowest redshift bin to $\sim 7\%$
and possibly higher.  Removal of the intrinsic alignment effect by the 
downweighting of nearby galaxy pairs will therefore be vital for SNAP. 
\end{abstract} 
 
 
\title{Intrinsic Galaxy Alignments: estimation and removal} 
\section{Introduction} 
The detection of weak gravitational lensing by large scale structure 
is a direct way to measure the total matter distribution in the 
Universe, demanding no assumptions for how luminous matter traces the 
dominant,  largely unknown, dark matter component.  Detected by 
several groups, weak gravitational lensing is now a well established 
technique, used successfully to set joint 
constraints on the matter density parameter $\Omega_m$ and the 
amplitude of the matter power spectrum, $\sigma_8$, 
\cite{Maoli,Rhodes,vWb01,HYG02,BMRE,Jarvis,MLB02,Hamana}, to measure the 
bias parameter $b$ \cite{HYG01,PenLu03}, and has recently been used to 
directly extract the 3D non-linear matter power spectrum 
$P_\delta(k)$ \cite{TegZald02,PenLu03}.   Combined 
with cosmic microwave background observations, weak lensing can 
provide strong constraints for $\sigma_8$ and $\Omega_m$ 
as the degeneracies in each measurement are almost 
orthogonal in the $\sigma_8 - \Omega_m$ plane \cite{MLB02,Contaldi}. 
 
Unlike many other tests of cosmology, 
weak lensing surveys with photometric redshift information can 
tightly constrain cosmological quintessence models and the 
equation of state parameter $w$, which  
will be key to our understanding of dark energy,
\cite{Heavens03,RefSNAP03,Benabed,JainTay}.  With the increased image 
resolution available from multi-colour space-based lensing surveys 
it will also be possible to construct high-resolution projected dark 
matter maps, and 3D dark matter maps of mass concentrations $> 1 
\times 10^{13} M_{\odot}$ \cite{AndyT,HuKeeton,BaconTay,MaseySNAP03}. 
With future deeper and wider multi-colour surveys, for example the 
Canada-France-Hawaii Telescope Legacy Survey (CFHTLS) ({\it 
www.cfht.hawaii.edu/Science/CFHTLS}) and the space-based
SuperNova/Acceleration  
Probe (SNAP) ({\it snap.lbl.gov}),  weak gravitational lensing will soon 
reach its `era of high precision cosmology', provided it can get a 
good handle on the many causes of systematic errors that arise when 
trying to detect this minute weak lensing signal. 
 
Sources of systematic errors in weak lensing analysis arise from the 
shearing and smearing of galaxy images caused by the atmosphere, 
telescope optics, and detectors (\pcite{KSB}; hereafter KSB; \pcite{LK97}). 
With excellent seeing 
observing conditions, or space-based data, combined with instruments 
and detectors that are designed with weak lensing detection in mind, 
these effects can be minimised and corrected for (see for example 
\pcite{RhodesSNAP03} and \pcite{BRE}).  Aside from these 
observational effects there is a potentially significant error arising 
from a key assumption made for all weak lensing studies, that galaxy 
ellipticities are randomly oriented on the sky. 
Gravitational interactions during galaxy 
formation could however produce intrinsic shape correlations between 
nearby galaxies, mimicking to an extent weak lensing shear 
correlations.  As the new generation of wide-field deep weak lensing 
surveys beat down their observational sources of systematics, it is 
this additional source of intrinsic ellipticity correlations that 
could limit the accuracy of cosmological parameter estimation from 
weak lensing studies.  The extent to which this is true will be aided 
by the study in this paper. 
 
Observational evidence for the existence of intrinsic galaxy 
alignments comes from the detection of galaxy ellipticity correlations 
in low redshift surveys where weak lensing shear correlations are 
negligible, for example in the SuperCOSMOS survey \cite{BTHD02}, and 
in the Tully catalogue \cite{LP02}.  Theoretically, the intrinsic 
alignment of nearby galaxies has been investigated through numerical 
simulations and analytical techniques.  These have provided estimates 
of the order 10\% contamination to weak lensing measurements from
surveys with median redshift $z_m \sim 1.0$ 
(\pcite{HRH00}, hereafter HRH, \pcite{CM00,CKB01,CNPT01}, hereafter 
CNPT; \pcite{LP01,HZ02,Porciani,Jing}, hereafter Jing; 
\pcite{Mackey02}).  Whilst there is broad agreement between these 
studies on the effect for 
weak lensing measurements, the finer details can differ by up to an order 
of magnitude or more, with the numerical simulations generally 
predicting a higher level of contamination than the semi-analytic 
studies.  In this paper, we re-examine this issue, and show that a 
combination of misalignment of the baryon and halo angular momentum, 
as determined by \scite{vdBosch02}, 
and the finite thickness of disk galaxies, modifies the predictions of 
HRH bringing them into good agreement with the semi-analytic model of 
CNPT. 
 
With redshift information, it has been shown that  
the intrinsic signal can be suppressed 
in weak lensing analysis by downweighting galaxy pairs which are 
physically close (\pcite{HH03,KingSch02}). This can be done 
optimally without significantly increasing the shot noise in the 
final weak lensing analysis, and for this it is helpful (although 
not necessary) to have a good estimate of the level of 
contamination.  Obtaining an observationally constrained estimate 
is therefore one of the purposes of this paper,  using three weak 
lensing surveys: a re-analysis of COMBO-17 \cite{MLB02}, and the
published results of the Red-sequence Cluster Survey (RCS; 
\pcite{HYGBHI}) and the VIRMOS-DESCART survey \cite{vWb02}. 

This paper is organised as follows.  In 
Section~\ref{sec:IAmodels} we review three different  
theoretical models for the intrinsic ellipticity correlations between 
nearby galaxies, taken from HRH, Jing and CNPT. 
We present a modification to the HRH analysis that we apply to
numerical simulations in Section~\ref{sec:newmodel},  
finding excellent agreement with the observed
intrinsic alignment signal from the SuperCOSMOS survey. 
Using the three different intrinsic 
alignment models: HRH, Jing, CNPT, and our modified HRH model which we
shall call HRH* hereafter, 
we then determine intrinsic alignment contributions to the aperture 
mass B mode statistic $M_\perp$ \cite{Sch98}, for the RCS and the 
VIRMOS-DESCART survey.   Using published measurements of 
$M_\perp$ as upper limits for the intrinsic galaxy alignment 
contribution, we show in Section~\ref{sec:MapB} that, assuming 
there is no evolution in galaxy clustering, we can reject the 
intrinsic alignment models of Jing and HRH.  In 
Section~\ref{sec:C-17}, we observationally constrain and remove 
the contribution to COMBO-17's weak lensing measurements by 
intrinsic galaxy alignments.  This is made possible due to the 
highly accurate photometric information of the COMBO-17 survey, 
through the application of an optimal intrinsic alignment 
contamination removal method as described in \scite{HH03}. In 
Section~\ref{sec:evo}, we investigate the effect that galaxy 
clustering evolution would have on our results.  We look at the 
implications for weak lensing analysis in 
Section~\ref{sec:implications}, constraining $\sigma_8$ and 
$\Omega_m$ with the shear correlation function statistic from 
COMBO-17, where we now include marginalisation over our 
measured intrinsic alignment signal.  We also determine estimates of
the contamination of shear correlation measurements 
from the CFHTLS and SNAP surveys, summarising and concluding in 
Section~\ref{sec:conc}. 
 
\section{Intrinsic ellipticity correlations} 
\label{sec:IAmodels} 
 
The ellipticity of a galaxy approximated as an ellipse with axial 
ratio $q$, at position angle $\phi$ measured counter-clockwise 
from the $x$ axis, can be defined as 
\be 
\left( 
\begin{array}{c} 
e_1 \\ 
e_2 
\end{array} 
\right) = 
\frac{q^2 - 1}{q^2 + 1} 
\left( 
\begin{array}{c} 
\cos 2\phi \\ 
\sin 2\phi 
\end{array} 
\right). \ee In the weak lensing r\'{e}gime the observed (complex) 
ellipticity $e$ is related to the source ellipticity $e^s$ and 
the complex shear $\gamma$ by $e \simeq e^s+2 \gamma$.  With many 
source galaxies the shear can be estimated from 
\begin{equation} 
\gamma \simeq \frac{1}{2} \langle e \rangle \equiv \frac{1}{2} 
\langle e_1 + i e_2 \rangle . 
\end{equation} 
See \scite{RRG00} for an exact relationship.
A measurement of the shear correlation function $\langle \gamma 
\gamma^* \rangle$ can be directly related to the matter power 
spectrum (see \pcite{Bible} and references therein), but to 
estimate it observationally includes an uncertain contribution 
from intrinsic correlations between the ellipticities of nearby 
galaxy pairs  $ I_{ab} \equiv e_a^s e_b^{s*}$ if, when averaging over many
galaxy pairs, $ \langle I_{ab} \rangle \neq 0 $.  Hence the expectation
of the shear correlation function is given by
\begin{equation} 
E[\gamma\gamma^*:\theta]  = 
\frac{1}{4} \sum_{ab} \left(e_a({\bf x}) e_b^*({\bf x} + 
\theta)-I_{ab}  \right). \label{LensingC} 
\end{equation} 
Several groups have found theoretical support for non-zero intrinsic 
correlations, and the existing methods used to determine its amplitude can be 
broadly split into two approaches.  The first employs numerical 
simulations with varying simplifying assumptions, relating galaxy 
shape to dark matter halo properties. The second uses analytic 
techniques, relating ellipticity correlations to initial linear 
tidal shear field correlations.  These methods approximate $I_{ab}$ as
a theoretically determined three-dimensional ellipticity
correlation function 
$\eta(r)$ which depends only on the comoving distance between the
galaxy pair $r \equiv |{\bf r}|$:
\be 
\eta(r) = 
\langle e({\bf x})e^*({\bf x}+{\bf r})\rangle. 
\ee 
We project $\eta(r)$ 
into two dimensions using a modified Limber equation, in order 
to compare it with observed angular ellipticity correlations.  In the
absence of weak gravitational lensing, for  
example with the low redshift SuperCOSMOS data \cite{BTHD02}, 
measured angular ellipticity correlations can be directly 
compared to those predicted from theoretical models: 
\be 
C_I(\theta) = \frac{\int dz_a dz_b   \phi_{z}(z_a) \phi_{z}(z_b) 
\left[1+\xi_{gg}(r_{ab})\right] \eta(r_{ab})}{ \int dz_a dz_b 
\phi_{z}(z_a) \phi_{z}(z_b) \left[1+\xi_{gg}(r_{ab})\right]}, 
\label{eqn:Ctheta} \ee where $\phi_{z}(z_a)$ is either a known 
redshift distribution, for example the COMBO-17 photometric redshift 
distribution shown in Fig.~\ref{fig:C17zs} of the Appendix, or can be 
approximated by 
\begin{equation} 
\phi_z(z) \, \propto \, z^2 \exp \left[ 
-\left(\frac{z}{z_0}\right)^{1.5} \right], 
\label{eqn:selfunc} 
\end{equation} 
for a survey with median redshift $z_m \approx 1.4 z_0$. 
Most studies have found $\eta(r_{ab})$ to be
significantly non-zero only for 
galaxies closer than a few tens of Mpc, where the comoving separation, 
$r_{ab}$ is given accurately enough in a flat universe by 
\be 
r_{ab}^2 \simeq (w_a-w_b)^2 + \left( \frac{w_a+w_b}{2} \right)^2 \theta^2, 
\ee 
where $w$ is the comoving radial geodesic distance.  
The signal depends on the 
galaxy two-point correlation function, $\xi_{gg}(r)$, since it 
determines how many galaxy pairs observed at a given angular separation 
are physically close together.  We take the correlation function to be 
non-evolving (\pcite{Cole94,Giavalisco,Adelberger}): 
\be 
\xi_{gg}(r) = \left( \frac{r}{r_0} \right)^{\gamma}, 
\label{eqn:non-evo} 
\ee 
with $\gamma = -1.8$ and $r_0 = 5 h^{-1}{\rm Mpc}$. In 
Section~\ref{sec:evo} we will also consider the 
effect of redshift dependence in the galaxy correlation function, as claimed 
by \scite{lefevre,PLSO,Baugh,Carlberg00,Firth,Wilson}. 
 
\begin{figure} 
\centerline{\psfig{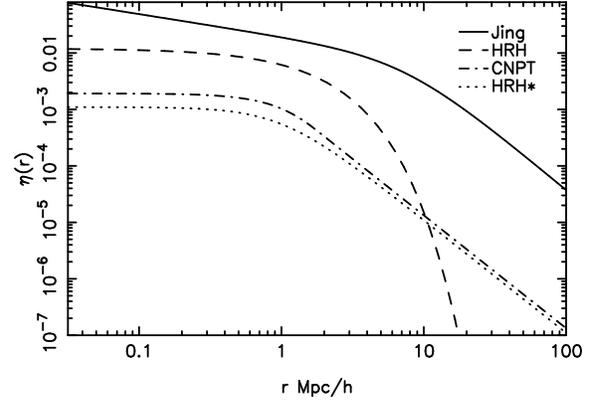}} 
\caption{Intrinsic ellipticity correlation models. Comparison of 
  different functional fits to three intrinsic alignment studies, 
  Jing (solid), HRH (dashed) and CNPT (dot dash).  The modified HRH 
  model that is measured and defined in Section~\ref{sec:newmodel} is also 
  plotted (dotted).} 
\label{fig:3Dcor} 
\end{figure}

In this Section we will focus on three studies: HRH and Jing, which 
are based on numerical simulations, and CNPT, 
where the strength of ellipticity correlations are 
analytically computed in linear theory.   
Fig.~\ref{fig:3Dcor} compares three published 
functional fits, defined in Section~\ref{sec:fits}, for the 
three-dimensional ellipticity correlation functions, $\eta(r)$, showing 
that whilst there is some rough agreement in the results from these studies, 
the numerical simulations in general, predict a stronger 
intrinsic alignment effect compared to the analytical technique. 
 
\subsection{Intrinsic Alignment Models} 
\label{sec:fits} 
Assuming that luminous matter forms galaxies in all dark matter halos 
above a minimum mass limit, $M_{h}$, it is possible to acquire a 
three-dimensional catalogue of galaxy shapes from a large N-body dark matter 
simulation.  HRH have modelled the shape of each luminous galaxy 
as an infinitesimally thin disc placed perpendicular to its angular 
momentum vector, which is assumed to be 
perfectly aligned with that of its parent dark matter halo.  This 
resulted in ellipticity correlations that were fitted with an 
exponential
\be 
\eta_{\rm HRH}(r) = 0.012 \exp\left(-\frac{r}{1.5 h^{-1} Mpc}\right). 
\label{eqn:HRH} 
\ee 
Using a higher resolution N-body simulation, Jing  
assumed that the 
galaxy ellipticity was equal to that of its parent dark matter halo, 
resulting in ellipticity correlations with a best fitting 
functional form defined for $\langle e_1 e_1 \rangle$, 
\be 
\eta_{\rm Jing}(r) \approx 2\langle e_1 e_1 \rangle = 2 \,\,\frac{3.6 
\times 10^{-2} \left( \frac{M_h}{10^{10}h^{-1}Mpc} 
\right)^{0.5}}{ 
r^{0.4}(7.5^{1.7} + r^{1.7})}, 
\ee 
where in this paper we will use $M_h = 6.9 \times 10^{11} M_{\odot}$, which 
corresponds to the minimum mass limit used by HRH.  
HRH found a similar shape for the ellipticity correlations of 
dark matter haloes.
CNPT related galaxy ellipticity correlations to the initial 
correlations in the tidal shear field.   
Assuming that the density field has a Gaussian distribution, CNPT found 
that the ellipticity correlation is well approximated at large $r\gs 3 
h^{-1}$Mpc by  
\be 
\eta_{\rm CNPT}(r) = \frac{a^2 
  \beta^2}{84}\frac{\varepsilon^2(r)}{\varepsilon^2(0)} , 
\label{eqn:CNPT} 
\ee 
where $a$ quantifies the uncertainty in the correlation between the shear 
field and a galaxy's moment of inertia, 
$\beta$ accounts for the non-zero thickness of galaxy disks where 
$e_{\rm gal} = \beta e_{\rm thin \, disc}$,  and 
$\varepsilon(r)$ is a density correlation function. 
For the purpose of this paper, we use an analytic solution to 
equation (\ref{eqn:CNPT}) for the simple model proposed in CNPT, who took
$a = 0.55$, inferred from numerical simulation results \cite{HRH00},
$\beta = 0.73$, 
from the mean of the measured ellipticity distribution of the APM
survey \cite{LML92} and from the observed distribution of
ellipticities measured for lensing studies \cite{Ebbels}, and  
$\varepsilon(r) \propto 1/r$, where we also follow CNPT by including
top-hat smoothing on $1h^{-1}{\rm Mpc}$ scales.  
 
\subsection{Intrinsic alignments of galaxies: revisiting theoretical
expectations}  
\label{sec:newmodel} 
 
\begin{figure} 
\centerline{\psfig{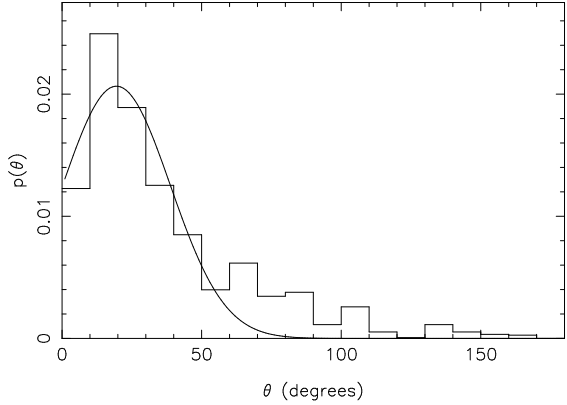}} 
\caption{Distribution of the misalignment angle between
the total angular momentum vectors of the dark matter and gas, as found
by van den Bosch et al. (2002).  Over-plotted is the best fit Gaussian
model with $\mu = 19.5^\circ$ and $\sigma = 19.3$.} 
\label{fig:VdB} 
\end{figure}

Recent results from numerical simulations of structure formation by
\scite{vdBosch02}, have shown that the angular momentum of
non-radiative gas and dark matter are actually poorly aligned with a
median misalignment angle of $\sim 30^\circ$.  Observational support
for some misalignment, albeit smaller, comes from strong lensing
considerations in the 
CASTLES survey \cite{Keeton98}.  With these results we are able to
advance the analysis of HRH, where instead of assuming perfect
alignment between the angular momentum of the halo and baryons, we
include a probability distribution for the misalignment.
Figure~\ref{fig:VdB} shows the probability distribution of the polar
misalignment angle $\theta$ as found by \scite{vdBosch02} and the best
fit Gaussian model.  \scite{vdBosch02} note that higher resolution
simulations are required to reduce the possible impact of discreteness
effects on this probability distribution, but they do conclude that
there is a true misalignment that is significantly different from
zero.

\begin{figure} 
\centerline{\psfig{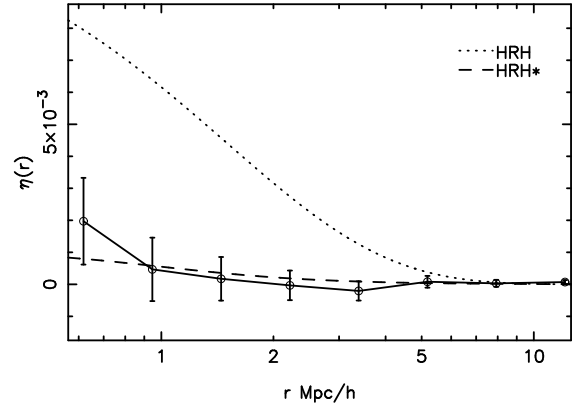}} 
\caption{3D ellipticity correlation function estimated from the 
$\Lambda$CDM Virgo simulation, assuming that the gas disc and dark 
matter halo are misaligned.  Over-plotted is the HRH model 
$\eta_{\rm HRH}(r)$ (dotted) and the best fit 
HRH modified model $\eta_{\rm HRH*}(r)$ (dashed).} 
\label{fig:HRHmod} 
\end{figure}

In light of these results we have reanalysed the N-body simulation
developed by the Virgo  
Consortium \cite{Jenkins}, for a $\Lambda$CDM cosmological model, 
($\Omega_m = 0.3$, $\Omega_\Lambda = 0.7$, $\sigma_8 = 0.9$, $\Gamma = 
0.2$),  approximating the probability distribution for the 
misalignment polar angle as found by \scite{vdBosch02}, 
as a Gaussian with mean $\mu = 20^\circ$ and width $\sigma =20^\circ$,
which is truncated at zero misalignment. 
We assign a random misalignment azimuthal angle around the original 
halo angular momentum vector.  The galaxy ellipticity is then determined 
from the angular momentum of the baryons, by,  
\be 
|e(L_z)| = 
\beta \left(\frac{1-L_z^2}{1+ L_z^2}\right), 
\label{eqn:betaellip} 
\ee 
where we now account for galaxy 
thickness, following CNPT by taking $\beta = 0.73$. 
Fig.~\ref{fig:HRHmod} compares the resulting 3D ellipticity 
correlation function $\eta(r)$, with the HRH model from 
equation (\ref{eqn:HRH}), (dashed), showing that with the inclusion of 
baryon halo angular momentum misalignment, the amplitude of the galaxy 
alignments predicted from numerical simulations is significantly reduced. 
We choose to fit a simple function 
\be
\eta(r) = \frac{A}{1 + (r/B)^2}.
\label{eqn:twopowlaw}
\ee  
We fix $B=1 h^{-1}$ Mpc, which is similar to the 
CNPT choice of a smoothing length of $1 h^{-1}$Mpc, and calculate the maximum
likelihood value for the amplitude $A$.  A small 
positive value is preferred, although as shown dot-dashed in 
Fig.~\ref{fig:likelihood}, it should be noted that the 
reduced signal from the misalignment means the data are actually 
consistent with zero.  The best fit shown dashed in Fig.~\ref{fig:HRHmod}
is ($r$ in $h^{-1}$ Mpc)
\be 
\eta_{\rm HRH*}(r) = \frac{0.0011}{1 + r^2}, 
\label{eqn:HRHmod} 
\ee 
which is consistent with the CNPT analytical model.  Note that the new
model fit lies above the old HRH model at large scales. This should
not be taken to imply that misalignment increases the correlations, 
rather, it reflects the fact that the old exponential fit of HRH
underestimated the (noisy) correlations on large scales.

\begin{figure} 
\centerline{\psfig{file=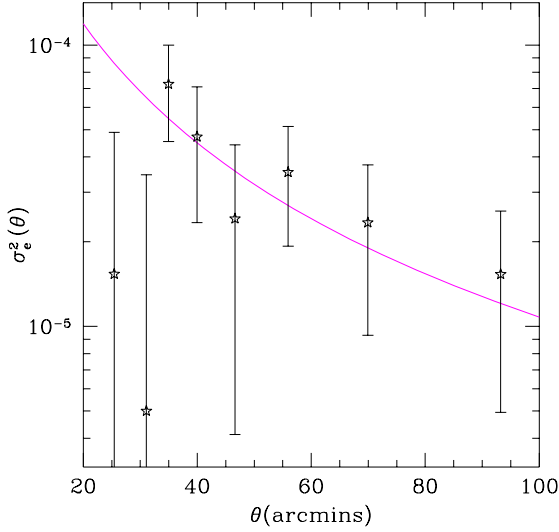,width=7.5cm,angle=0,clip=}} 
\caption{Comparison of the measured ellipticity variance in the 
  SuperCOSMOS survey, see Brown et al. (2002) for details, compared to the 
  ellipticity variance expected at this redshift from the modified HRH 
  model, $\rm{HRH*}$. } 
\label{fig:SuperC} 
\end{figure}

Comparing our HRH* model with the ellipticity 
variance measured in the low redshift SuperCOSMOS survey \cite{BTHD02},  
we find excellent agreement, especially on scales $>35$ arcmin, see
Fig.~\ref{fig:SuperC}.  We have calculated the HRH* 
angular ellipticity correlation function from 
equation (\ref{eqn:Ctheta}) with a median redshift $z_m = 
0.1$, and related it to the ellipticity variance statistic 
following \scite{BMRE}: 
\be 
\sigma^2_{e} (\theta)_{\,\Box} = \frac{2\sqrt{\pi}}{\theta^2} 
\int_0^{\theta} \, d\theta^\prime \, \theta^\prime C_I(\theta^\prime), 
\ee
where the factor $\sqrt{\pi}$ is a good approximation to use in order 
to scale the ellipticity 
variance measured in circular apertures of radius $\theta$ to the 
ellipticity variance as measured in SuperCOSMOS, in square cells of
length $\theta$ \cite{BRE}.  We calculate the likelihood of
the HRH* model from the SuperCOSMOS results
on scales $\theta > 20$ arcmin, allowing the amplitude to vary, see
Fig.~\ref{fig:likelihood}.  Below this angular scale the SuperCOSMOS results 
are extremely noisy and could suffer contamination from residual point
spread function anisotropy distortions, (see \scite{BTHD02} for details).  
We therefore omit these points from the likelihood fitting procedure,
finding the best fit amplitude $A = 0.0009 \pm 0.0005$, (95\%
confidence limits).
 
\section{Upper limits on the intrinsic alignment signal: Aperture mass B modes} 
\label{sec:MapB} 
 
A good diagnostic for determining the level of systematic errors 
present in weak lensing measurements, is to decompose the 
shear correlation signal into E and B modes. This was first proposed 
by \scite{CNPT02}, and is now a standard statistical test for the 
presence of non-lensing contributions to weak lensing measurements. 
Weak gravitational lensing produces curl-free distortions (E-type), 
and contributes only to the B-type distortions at small angular 
scales, $\theta < 1$ arcmin, 
due to source redshift clustering \cite{SchvWM02}. 
A significant detection of a B-type signal in weak lensing surveys is 
therefore an indication that ellipticity correlations exist either 
from residual systematics within the data and/or from intrinsic galaxy 
alignments. 
 
The decomposition has previously been carried out by determining the B mode 
shear power spectrum, $C_l^{\beta\beta}$ \cite{MLB02,PenWM} or by
using the aperture mass statistic, $M_{ap}$
\cite{HYG02,vWb02,Hamana,Jarvis}, where the aperture mass statistic is
defined to be the aperture weighted dimensionless surface mass density
of a lens, $\kappa$ \cite{Kaiser94}, such that 
\begin{equation}
  M_{\rm ap}(\theta) = \int_0^\theta d^2\vartheta\,U(\vartheta)\,
  \kappa(\bm{\vartheta})\;,
\end{equation}
where the simplest radial filter function $U(\vartheta)$ is given by,
\begin{equation}
  U(\vartheta) =
  \frac{9}{\pi\theta^2}\,\left(1-\frac{\vartheta^2}{\theta^2}\right)\,
  \left(\frac{1}{3}-\frac{\vartheta^2}{\theta^2}\right),
\end{equation}
\cite{Sch98}.  With this choice of filter function the aperture mass
$M_{ap}$ is related to the power spectrum by  
\be 
  \langle M_{\rm ap}^2\rangle(\theta) = 
  \frac{288}{\pi \theta^4}\,\int_0^\infty \frac {dk}{k^3} \, 
  P_\kappa(k)\,[J_4(k\theta)]^2\:, 
\ee 
where ${\rm J}_4(k\theta)$ is the fourth-order Bessel function of the 
first kind,  and 
$P_\kappa(k)$ is the convergence power spectrum at wave number $k$, 
related to the non-linear mass power spectrum $P_{\delta}$ by 
\begin{equation} 
P_\kappa(k) = \frac{9 H_0^4 \Omega_m^2}{4c^4} \int_0^{w_H} dw \, 
\frac{g^2(w)}{a^2(w)} \, P_\delta \left( \frac{k}{f_K(w)},w \right), 
\end{equation} 
$a(w)$ is the dimensionless scale factor, $H_0$ is the Hubble parameter and 
$\Omega_m$ the matter density parameter.  The second argument of $P_\delta$ 
allows for time-evolution of the power spectrum.  $g(w)$ is a 
weighting function locating the lensed sources, 
\begin{equation} 
g(w) = \int_w^{w_H}\, dw'\ \phi(w') 
\frac{f_K(w'-w)}{f_K(w')}. 
\label{eqn:W} 
\end{equation} 
$\phi(w(z))dw$ is the observed number of galaxies in $dw$, and
$w_H$ is the horizon distance \cite{Sch98}.  
 
In this paper we will focus on the aperture mass decomposition, which can be 
directly calculated from angular ellipticity correlation functions as follows: 
\be 
{\rm E:}\,\,\, \langle M_{ap}^2 \rangle = \int_{0}^{\infty} 
  \frac{d\vartheta \, 
  \vartheta}{2\theta^2} \left[ \xi_{+} (\vartheta) T_{+} 
  \left(\frac{\vartheta}{\theta}\right) + \xi_{-} (\vartheta) T_{-} 
  \left(\frac{\vartheta}{\theta}\right) \right] \nn 
\ee 
\be 
{\rm B:}\,\,\, \langle M_{\perp}^2 \rangle = \int_{0}^{\infty} 
  \frac{d\vartheta \, 
  \vartheta}{2\theta^2} \left[ \xi_{+} (\vartheta) T_{+} 
  \left(\frac{\vartheta}{\theta}\right) - \xi_{-} (\vartheta) T_{-} 
  \left(\frac{\vartheta}{\theta}\right) \right] 
\label{eqn:MapB} 
\ee 
where, 
\be 
\xi_{\pm}(\vartheta) = \langle \gamma_t \gamma_t \rangle_{\vartheta} 
\pm \langle \gamma_r \gamma_r \rangle_{\vartheta} 
\approx \frac{1}{4} \left( \langle e_t e_t \rangle_{\vartheta}   \pm 
\langle e_r e_r \rangle_{\vartheta}  \right), 
\ee 
and $T_{\pm}$ are formally given in \scite{CNPT02}. 
The tangential and radial 
ellipticity parameters $e_t$ and $e_r$ are defined such that: 
\be 
\left( 
\begin{array}{c} 
e_t \\ 
e_r 
\end{array} 
\right) = 
\left( 
\begin{array}{cc} 
\cos 2\phi & \sin 2\phi  \\ 
-\sin 2\phi & \cos 2\phi 
\end{array} 
\right) 
\left( 
\begin{array}{c} 
e_1 \\ 
e_2 
\end{array} 
\right), \ee 
where $\phi$ is defined to be the 
angle between the $x$ axis and the line joining a galaxy pair. 

In the absence of contaminating 
non-lensing sources we expect $M_{\perp}\approx0$, but this is found 
not to be the 
case in all the weak lensing surveys to date that have measured this 
diagnostic statistic.  We therefore calculate the contribution to the 
aperture mass B mode from the four intrinsic alignment models defined 
in Sections~\ref{sec:fits} and~\ref{sec:newmodel}.  This is calculated 
for two different depth weak lensing surveys, the 
RCS with a median redshift $z_m \approx 0.56$, 
\cite{HYGBHI}, and the VIRMOS-DESCART survey with $z_m \approx 1.0$, 
\cite{vWb02}. 
 
\begin{figure} 
\begin{tabular}{c} 
\psfig{file=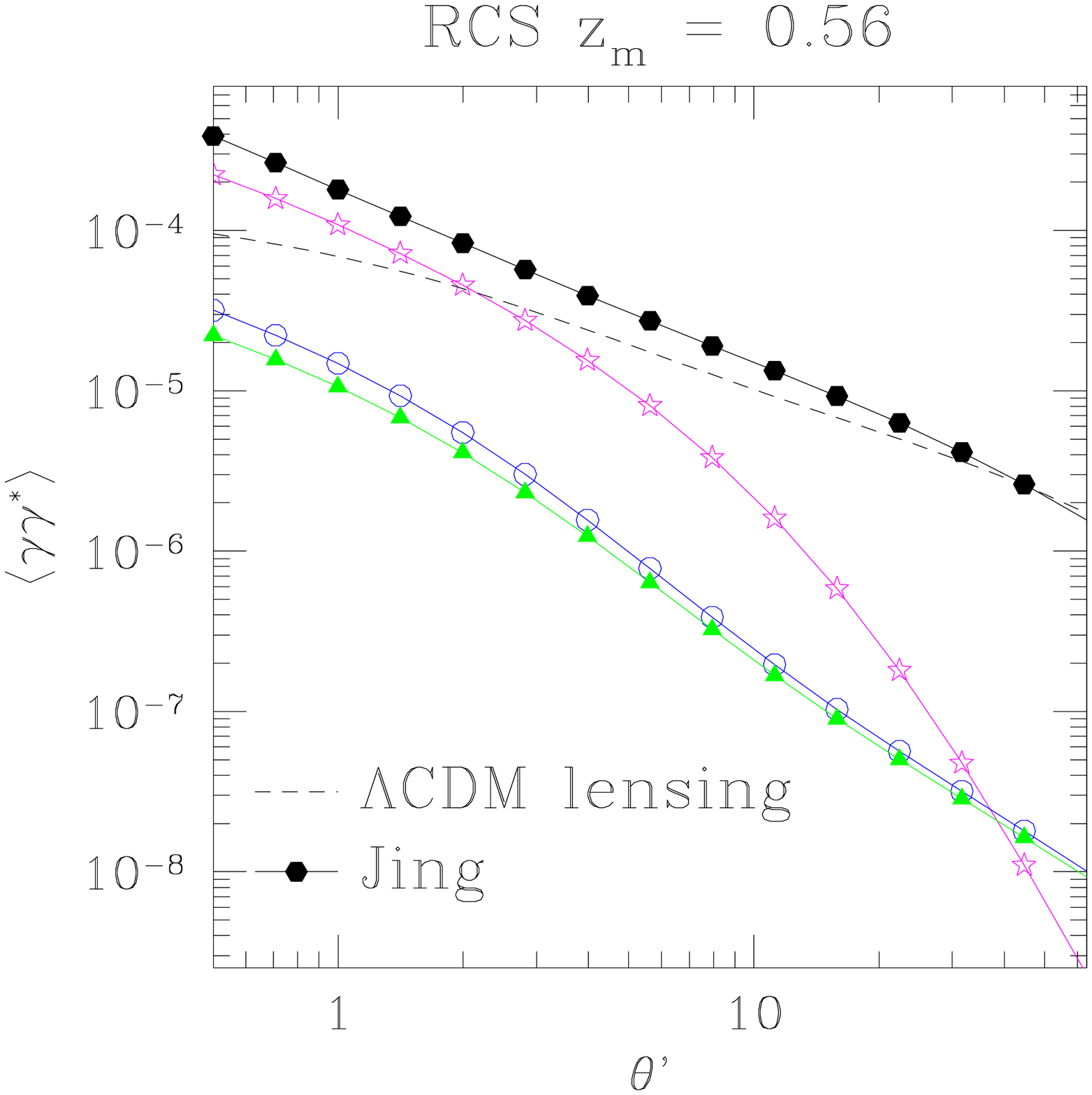,width=7.0cm,angle=0,clip=} \\ 
\psfig{file=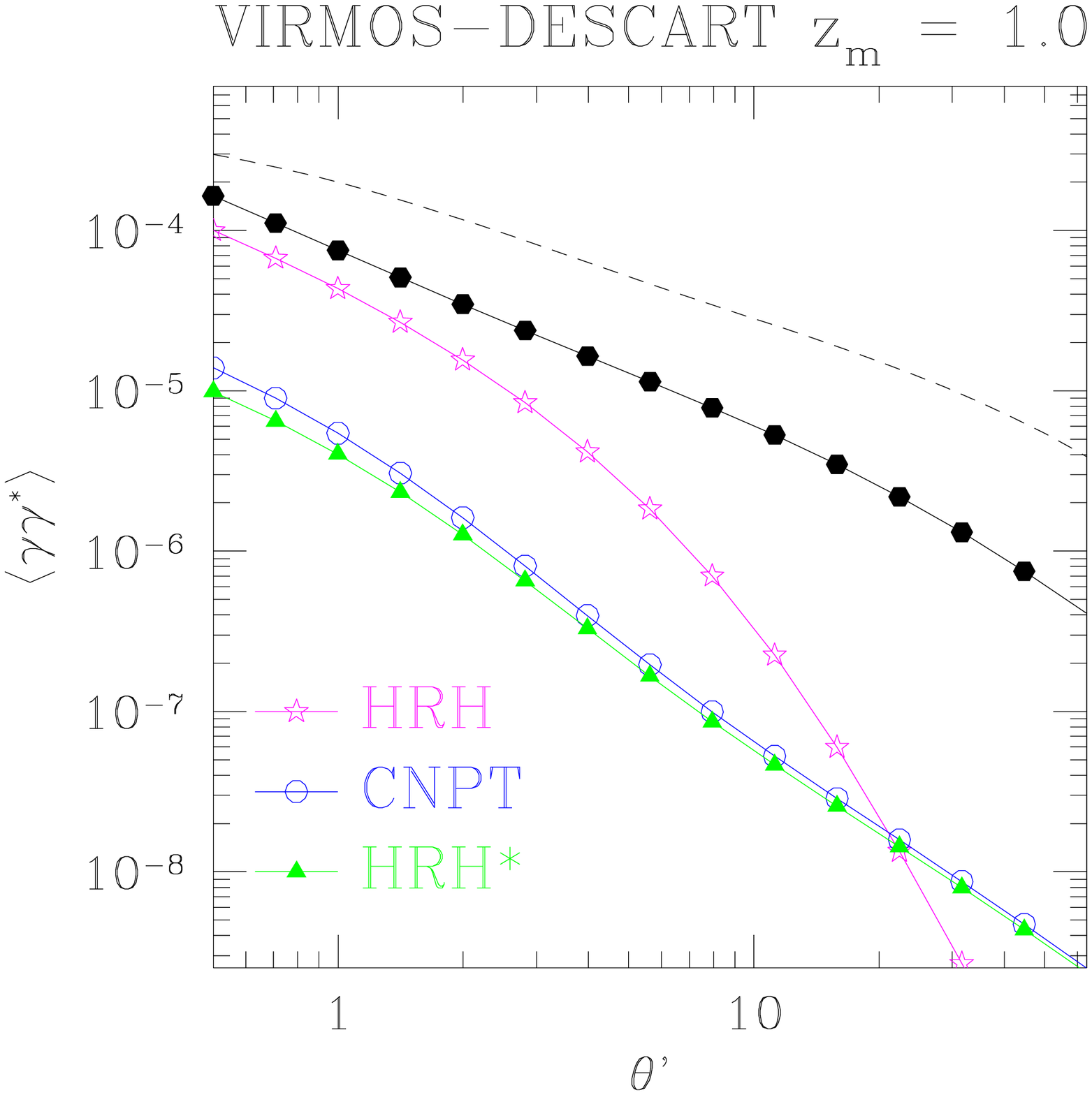,width=7.0cm,angle=0,clip=} 
\end{tabular} 
\caption{Predicted intrinsic alignment angular shear correlation 
functions $\langle \gamma \gamma^* \rangle_I$,  Jing (filled), HRH (stars), CNPT (circles) and HRH* (triangles). 
The intrinsic alignment contributions are compared to the angular 
shear correlations from $\Lambda$CDM weak gravitational lensing, 
normalised to $\sigma_8 = 0.8$, (dashed), 
for the RCS with depth $z_m = 0.56$, 
(upper) and for the VIRMOS-DESCART survey with depth $z_m = 1.0$, (lower).} 
\label{fig:Ctheta} 
\end{figure}

Fig.~\ref{fig:Ctheta} shows the predicted angular shear correlation function, 
$\langle \gamma \gamma^* \rangle (\vartheta)_I = \frac{1}{4}
C_I(\vartheta)$, for each 
intrinsic alignment model calculated for  
both survey depths.   We also plot for comparison 
the shear correlations expected from weak gravitational lensing, 
for an $\Omega_{m} = 0.3$, $\Omega_{\Lambda} = 0.7$ cosmology with a 
$\Gamma = 0.2 $, $\sigma_8 = 0.8$, nonlinear CDM matter power 
spectrum, $P_\delta$ calculated using the fitting formula from \scite{Rob}: 
\begin{equation} 
\langle \gamma \gamma^* \rangle(\vartheta)_{\rm WGL} = 
\frac{1}{2\pi}\int dk \,k \,P_\kappa(k) \, J_0(k\vartheta). 
\label{eqn:gg} 
\end{equation} 
 
\begin{figure} 
\centerline{\psfig{file=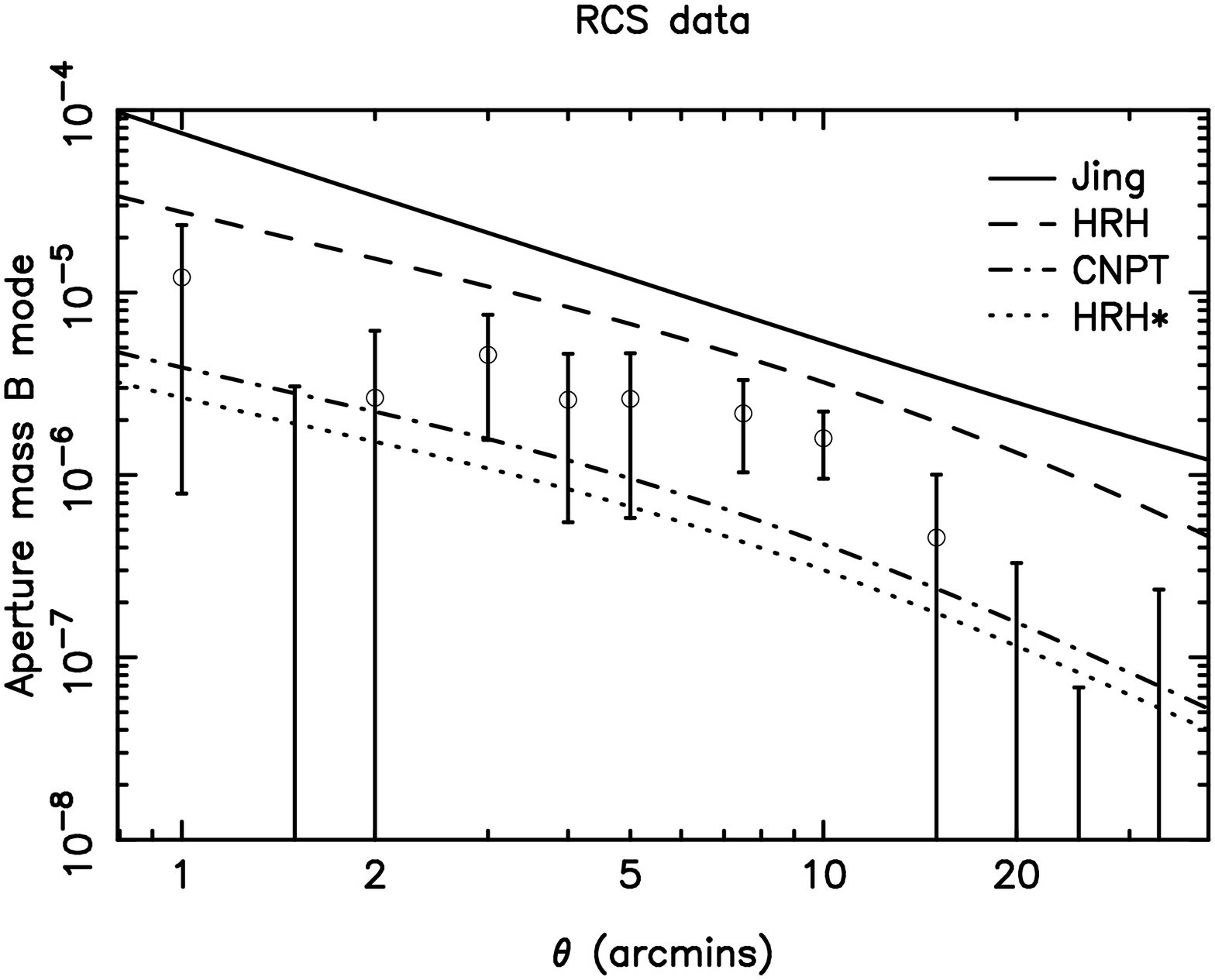,width=7.0cm,angle=270,clip=}} 
\centerline{\psfig{file=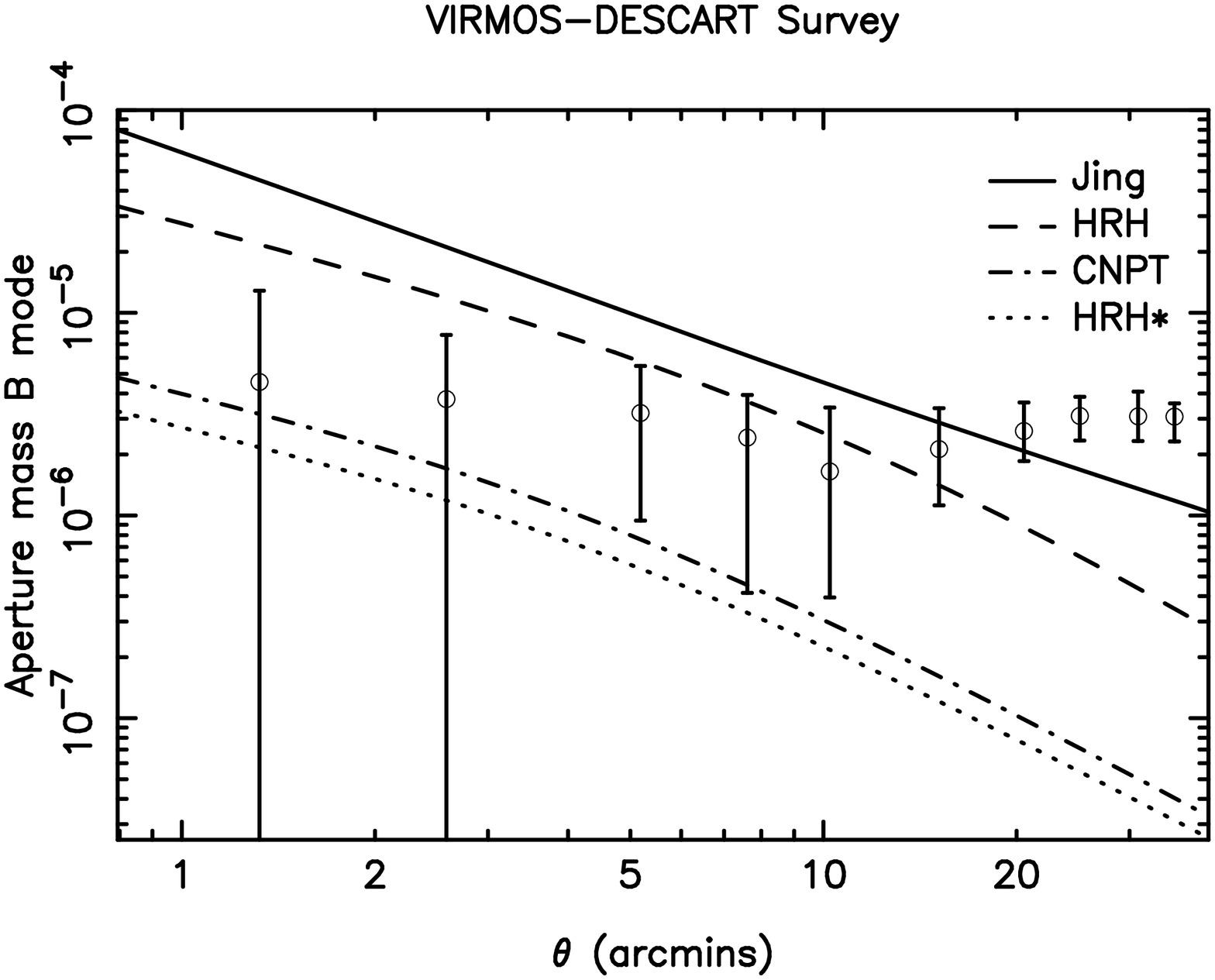,width=7.0cm,angle=270,clip=}} 
\caption{Intrinsic alignment contributions to the aperture mass B mode 
  $\langle M^2_\perp \rangle$; 
  Jing(solid), HRH (dashed), CNPT(dot dash) and HRH* (dotted). The upper panel 
  compares intrinsic alignment predictions for the RCS 
  survey with their measured B mode.  The lower panel compares intrinsic 
  alignment predictions for the VIRMOS-DESCART with their measured B mode.} 
\label{fig:M_ap} 
\end{figure}

We calculate the aperture mass B mode contributions from the 
shear correlation functions shown in Fig.~\ref{fig:Ctheta}, 
through equation (\ref{eqn:MapB}). 
CNPT find that $\langle e_t e_t\rangle_\vartheta \approx 
\langle e_r e_r\rangle_\vartheta$ at small angular scales, deviating
slightly at separations $\vartheta>3$ arcmin, and this is consistent with the 
results we find.  We therefore make the simplifying 
assumption that $\langle e_t e_t\rangle_\vartheta = 
\langle e_r e_r\rangle_\vartheta$ at all angular scales, i.e that
there is no preferred intrinsic tangential or 
radial alignment of galaxy ellipticities, and hence $\xi_{-} = 0$ such that 
\be 
\langle M^2_{\perp \, ({\rm IA})} (\theta) \rangle = \frac{1}{2}
\int_{0}^{\infty}  
  \frac{d\vartheta \, 
  \vartheta}{\theta^2} \left[ \xi_{+} (\vartheta) T_{+} \left( 
  \frac{\vartheta}{\theta}\right) \right], 
\ee 
where $\xi_{+}(\vartheta) = \frac{1}{4}C_I(\vartheta)$ and we use 
the following analytic expression 
for $T_{+}(x)$, derived by \scite{SchvWM02}, which vanishes at $x>2$: 
\ba 
T_{+}(x)&=&\frac{6(12-15x^2)}{5} \left[ 1 - 
  \frac{2}{\pi}\arcsin\frac{x}{2}  \right]  + \frac{x \sqrt{4 - 
  x^2}}{100\pi}\nn 
& & \times \left(110+2320x^2-754x^4+132x^6-9x^2 \right). 
\label{Tplus} 
\ea 
Fig.~\ref{fig:M_ap} shows the aperture mass B mode measurements reported 
by the RCS \cite{HYG02}, (upper panel), and VIRMOS-DESCART survey, 
\cite{vWb02}, (lower panel), 
compared to the contributions from the four different intrinsic 
alignment models.  The observed 
B modes can be attributed in part to residual systematics remaining in
the data and in part to intrinsic galaxy alignments.
Whilst most models for intrinsic galaxy alignments predict B modes
there are alternatives such as the model proposed by \scite{CKB01} which
does not.  Known data 
systematics introduce positive correlations and we therefore 
consider these observations as upper limits for intrinsic alignment B modes 
within these surveys.  This is probably a valid assumption on small scales
over which the distorting point spread function (PSF) is relatively
constant, provided a consistent subtraction method for  
the PSF has been applied.   Highly anisotropic PSF distortions which
vary from $e^*_i \rightarrow -e^*_i$ across the image could introduce
a negative aperture mass B mode on large scales. This could in 
principle, introduce a systematic negative aperture mass
B mode at large scales $\theta>10$ arcmin, preventing the use of B modes
as an upper limit for intrinsic alignment contamination at these scales.

With no evolution in galaxy clustering, 
considering the observed B modes as upper limits for $\theta<10$
arcminutes, we find that the Jing and HRH models are strongly rejected by 
the RCS results and that the CNPT and HRH* models are favoured by both
the VIRMOS-DESCART and the RCS.  We note that none of the 
intrinsic alignment models can account for the significant B mode at large 
angular scales $\theta>15$ arcmin found in the VIRMOS-DESCART data, which 
may be caused by other data systematics \cite{Hoekstra03}. 
 
\section{Estimation and removal of the intrinsic alignment signal: 
  shear correlations}  
\label{sec:C-17} 
 
Decomposing cosmic shear correlation measurements into E and B modes 
provides an estimate of the intrinsic alignment contamination in weak lensing 
measurements.  However it is not clear how to correct the E mode given 
a measured non-zero B mode. To date the practice has been to add in 
quadrature the B mode to the E mode error \cite{HYG02}. 
An alternative technique, applicable to weak lensing surveys with 
photometric redshift estimates, is to remove, essentially completely, 
the intrinsic alignment signal by downweighting pairs of galaxies 
which are likely to be physically close (\pcite{KingSch02,HH03}, 
hereafter HH).  \scite{KingSch03} have 
also proposed separating cosmic shear from intrinsic 
galaxy alignments using correlation function tomography. 
 
In this Section, we extend the method of HH, to determine the 
intrinsic alignment signal, as well as to remove it from the shear 
correlation function.  The method is reviewed in the appendix; 
close galaxy pairs are ignored in the shear correlation analysis, but 
in the optimised case, no more are removed than is strictly necessary, to 
avoid needlessly increasing the shot noise. 
 
\subsection{Constraining intrinsic galaxy alignments with COMBO-17: Method} 
\label{sec:constrainIA} 
 
COMBO-17 is a deep multi-colour wide-field optical survey carried out
with the Wide-field Imager (WFI) at the MPG/ESO 2.2m telescope
that spans 1.25 square degrees to limiting magnitude
${\it R}=25.5$ in five separate regions \cite{C17}.  
Multi-colour observations through a total of
17 filters, 5 broad-band ($\it{UBVRI}$) and 12 narrow band filters ranging 
from 420 to 914 nm, specifically chosen to facilitate accurate 
photometric redshift estimation with errors $\Delta_z \le 0.05$ 
reliable to $R<24$ currently span 1 square degree of the survey area.
The WFI instrument consists of a $ 4 \times 2 $  
array of $2048 \times 4096$ CCDs with pixel scale $0.238''$. 
Weak lensing studies have been carried 
out on the deep R-band images observed during the best seeing 
conditions \cite{MLB02}, and it is a subset of this R-band selected 
galaxy sample, limited to $R<24$ with a median redshift $z_m \sim 
0.6$, that we will use in this analysis. 
The selected survey area totals 0.75 square degrees in the A901, CDFS 
and S11 fields, yielding a catalogue of $3.55 \times 10^4 $ 
galaxies with KSB shape measurements and 
photometric redshifts accurate to $\Delta_z = 0.042$, (see 
\pcite{MLB02} for further details). 
 
\begin{figure} 
\centerline{\psfig{file=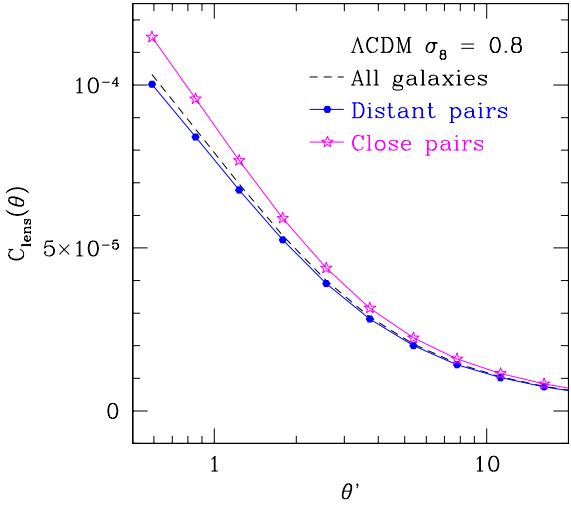,width=8.0cm,angle=0,clip=}} 
\caption{Predictions for the weak lensing shear correlation functions for 
the distant pairs and for the close pairs as described in the 
text. The difference in amplitude between the predicted signals is 
$\Delta C_{\rm lens}(\theta) \le 1.5\times 10^{-5}$ at all measured 
scales.} 
\label{fig:lens_models} 
\end{figure}

The total observed shear correlation function in the COMBO-17 dataset 
can be expressed as 
\be 
C(\theta)=C_{\rm lens}(\theta)+C_{\rm I}(\theta)+C_{\rm sys}(\theta), 
\ee 
where $C_{\rm lens}(\theta)$ is the cosmic shear weak lensing signal
$\langle \gamma \gamma^* \rangle$ equation (\ref{eqn:gg}),
and $C_{\rm I}(\theta)$ is the signal due to the intrinsic alignment 
of galaxy shapes equation (\ref{eqn:Ctheta}).  $C_{\rm sys}(\theta)$ is
the correlation function  
due to any observational systematic effects which may be present in 
the dataset (see \pcite{MLB02} for a robust estimate of systematic 
errors in the data).  $C_{\rm I}(\theta)$ can effectively be 
eliminated by excluding galaxy pairs which are closer than 
$\alpha(\theta)\Delta_z$ in redshift, where $\Delta_z=0.042$ is the 
typical redshift error for the COMBO-17 galaxies and $\alpha(\theta)$ 
is optimised to minimise the total error from intrinsic alignments and 
shot noise (see appendix).  By measuring 
the shear correlation function for pairs of galaxies which are sufficiently distant 
from each other, we exclude intrinsic alignments. Furthermore, if we 
also measure the shear correlations for only the close galaxy pairs 
which have been excluded from the analysis, we have two independent 
correlation functions from which we can estimate the intrinsic 
alignment signal from the close galaxy pairs:
\be 
C_{\rm IA}(\theta) = C_{\rm close}(\theta)-C_{\rm distant}(\theta) - 
\Delta C_{\rm lens}(\theta), 
\label{combo_ia} 
\ee 
where $C_{\rm close}(\theta)$ is the correlation function measured 
from the close pairs, $C_{\rm distant}(\theta)$ is the correlation 
function for the distant pairs and $\Delta C_{\rm lens}(\theta)$ is 
the difference in the correlations caused by weak gravitational  
lensing for the distant and 
close pairs.  Equation (\ref{combo_ia}) assumes 
that in the distant data set the intrinsic 
alignment signal is zero and that $C_{\rm sys}(\theta)$ is the same as for 
the close pairs.  To ensure all intrinsic alignment contamination has 
been removed from the distant dataset, we choose the most conservative 
estimate of $\eta_{\rm Jing}(r)$ in our HH weighting scheme. The 
assumption for $C_{\rm sys}(\theta)$ is reasonable since there is no 
reason to believe that systematic effects would depend on the redshift 
separation of galaxy pairs. To calculate $\Delta C_{\rm 
lens}(\theta)$, in Fig.~\ref{fig:lens_models} we plot the expected 
lensing signal for a median redshift $z_m = 0.6$ $\Lambda$CDM model, 
normalised to $\sigma_8 = 0.8$, for both the distant-pair dataset and for 
the close pairs, according to equation (A13) in HH. Although the difference is 
small, at the $10^{-5}$ level, and could be ignored,  
we will include it in our analysis assuming that $\Delta C_{\rm 
  lens}(\theta)$ is approximately the same for each field.  The weak lensing 
signal $C_{\rm lens}(\theta)$ will however  
differ significantly between the three fields, A901, CDFS and 
S11, due to the significant mass concentrations in two of the 
fields. The S11 field contains a fairly large cluster (Abell 1364) at 
a redshift of $z=0.11$ while the A901 field includes a supercluster 
system (Abell 901/902) at $z=0.16$.  In the following analysis, we 
have therefore applied equation (\ref{combo_ia}) to the three fields 
individually and combined the three resulting measurements of $C_{\rm 
IA}(\theta)$ with minimum variance weighting, where the weight $w_i$ for
each measurement with associated error $\sigma_i$ is given by $w_i =
1/\sigma_i^2$. This ensures that 
$C_{\rm lens}(\theta)$ is eliminated in the subtraction.  Note also 
that, for the various $C(\theta)$ calculations, we have removed 
cluster member galaxies from the A901 and S11 fields to eliminate 
contamination of the field intrinsic alignment signal by intra-cluster 
galaxy alignments \cite{Plionis03}. 
 
To estimate $C_{\rm close}(\theta)$ and $C_{\rm distant}(\theta)$ for 
the three fields, we follow the analysis described in 
\scite{MLB02}. We split each field into eight chip-sized sections and
calculate the relevant correlation function ($c_{\rm i,s}$) for each 
section, 
\be 
c_{\rm i,s}(\theta) = \frac{ \sum_{\rm ab} W_{\rm ab}\, 
\gamma(\bm{\theta}_a) \, \gamma^*(\bm{\theta}_b)} 
{\sum_{\rm ab} W_{\rm ab}}, 
\label{eqn:weightedcor} 
\ee 
where $W_{\rm ab}$ is the weight for the galaxy pair $ab$.  For the 
distant-pair data set $i=$ `distant', $W_{\rm ab} = 0$ if $ | \hat
z_a - \hat z_b| < \alpha  \Delta_z $ and $W_{\rm ab} = 1$ otherwise.
For the close pairs $i=$ `close' the  
weighting is simply reversed.  Our estimate of the  correlation 
function for the entire field is the average, 
\be 
C_i(\theta)=\frac{1}{N_s}\sum^{N_s}_{s=1} c_{i,s}(\theta), 
\label{corr} 
\ee 
where $N_s$ is the number of sections in a field. Note that here 
we are making the approximation that sections of the same field are 
uncorrelated. Neighbouring sections of the same field are, in fact, 
correlated but our approximation will be valid on scales 
which are small compared to the size of the sections  ($=8\times16$ 
arcmin). With this approximation, an estimate of the covariance matrix 
of the  $C_i(\theta)$ measurements is given by 
\ba 
&& {\rm cov}[C_i(\theta) C_j(\theta')] \simeq \nn && \,\,\,\,\,\,\,\, 
\frac{1}{N^2_s}\sum_{s=1}^{N_s} [c_{i,s}(\theta)-C_i(\theta)] 
[c_{j,s}(\theta')-C_j(\theta')]. 
\label{covcorr} 
\ea 
We have investigated the sensitivity of the results to the choice of
the number of sections each field is divided into, considering 4, 8
and 16 sections.  The most probable solution for
$\sigma_8(\Omega_m/0.27)^{0.6}$ varies by about 0.03, much less than the
statistical error, which is about 0.1 (see Section 6).  We do, however, add
this error in quadrature in the quoted results.
 
\subsection{An observationally constrained intrinsic alignment model} 
\label{sec:C17model} 
 
\begin{figure} 
\centerline{\psfig{file=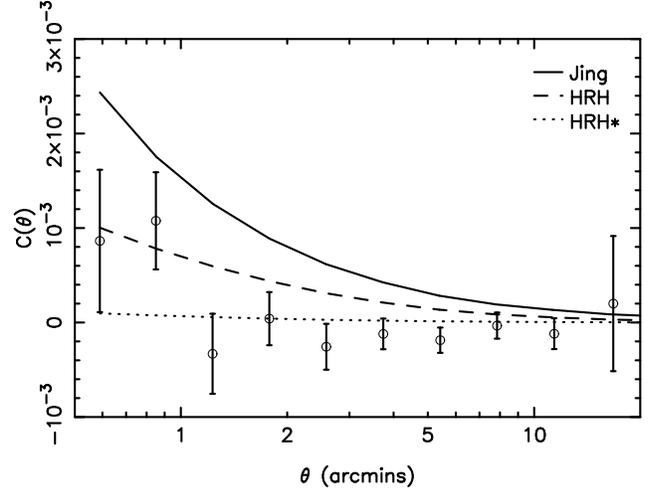,width=7.0cm,angle=270,clip=}} 
\caption{The intrinsic shear correlation functions from close galaxy pairs 
  in the COMBO-17 survey, 
  compared to predictions from Jing (solid), HRH (dashed) and HRH* 
  (dotted). The CNPT prediction lies slightly above the HRH* prediction 
  but on this scaling is indistinguishable and is therefore not plotted.} 
\label{fig:expairs} 
\end{figure}

\begin{figure} 
\vspace{-4cm} 
\centerline{\psfig{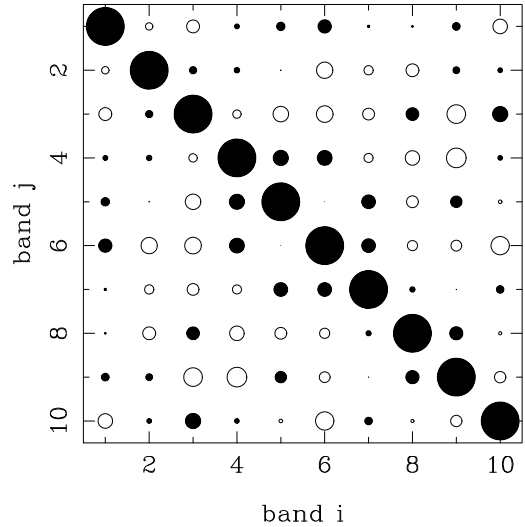}} 
\vspace{-3.5cm} 
\caption{Correlation matrix of the optimally combined $C_{\rm 
IA}(\theta)$ measurements plotted in Fig.~\ref{fig:expairs}. The area 
of each circle is proportional to the degree of  correlation between 
points $i$ and $j$. Filled circles denote that the points are 
correlated  whereas unfilled circles denote an anti-correlation 
between the points. The points are the  same as those plotted in 
Fig.~\ref{fig:expairs}, numbered 1 to 10, in order from left  to 
right.} 
\label{fig:covar} 
\end{figure}

We have applied equations (\ref{combo_ia}), (\ref{corr}) and (\ref{covcorr}) 
to the three fields, A901, CDFS and S11 and have 
combined the resulting measurements with minimum variance weighting 
to yield our final 
measurements of the intrinsic alignment signal for the close pairs 
in the COMBO-17 dataset. These final measurements are shown in 
Fig.~\ref{fig:expairs} along with the predicted intrinsic alignment 
signal for the close pairs, 
\be 
C_{IA}(\theta) = \frac{\int dz_a dz_b   \phi_{z}(z_a) 
\phi_{z}(z_b)  \left[1+\xi_{gg}\right]\langle W_{ab}\rangle 
\eta(r_{ab})}{ \int dz_a dz_b   \phi_{z}(z_a) \phi_{z}(z_b) 
\left[1+\xi_{gg}\right]\langle W_{ab}\rangle} 
\label{eqn:CIAtheta} 
\ee 
where we use three different models for the intrinsic alignment
$\eta(r_{ab})$, and $\langle W_{ab}\rangle$ is given by 
\be
\langle
W_{ab} \rangle = 
\frac{1}{2 \sqrt{\pi}} \int_{-\infty}^{\infty} dx \, \emph{e}^{-x^2}
\left[ {\rm erf} (y + x) -  {\rm erf} (v + x) 
\right]
\label{eqn:Wab}
\ee
where,
\be 
x\equiv \frac{\hat z_a - z_a}{ \sqrt{2} \, \Delta_z} \,\,\,\, y \equiv
\frac{z_a - z_b + \alpha\Delta_z }{\sqrt{2} \, \Delta_z} \,\,\,\, v
\equiv \frac{z_a - z_b - \alpha\Delta_z }{\sqrt{2} \, \Delta_z}.
\ee
\cite{HH03}. Fig.~\ref{fig:covar} shows the correlation matrix of the
$C_{\rm IA}(\theta)$ measurements, defined by  
\be 
 {\rm cor}(\theta,\theta')=\frac{ {\rm cov}(\theta,\theta') } 
 {\sqrt{{\rm cov}(\theta,\theta) {\rm cov}(\theta',\theta')}}, 
 \label{corr_matrix} 
\ee 
where ${\rm cov}(\theta,\theta')={\rm cov}[C_{\rm IA}(\theta),C_{\rm 
IA}(\theta')]$ is the covariance matrix of the $C_{\rm IA}(\theta)$ 
measurements, which we estimate from the data using 
equations (\ref{covcorr}) and (\ref{combo_ia}). 
One can immediately see from 
Fig.~\ref{fig:expairs}, in agreement with the aperture mass B mode 
analysis in Section \ref{sec:MapB}, that the measured 
COMBO-17 $C_{\rm IA}(\theta)$ strongly rejects the Jing model for the 
intrinsic alignment 
signal while the HRH model is also highly inconsistent with the 
data. The modified HRH model, HRH*, introduced in Section 
\ref{sec:newmodel}, and therefore also the CNPT model, 
are much better fits to the 
data. We note also  that the $C_{\rm IA}(\theta)$ measurements are 
consistent with a null result on scales $> 1$ arcmin although, from 
Fig.~\ref{fig:covar} it's clear that there are some (anti-) 
correlations  between the correlation function measurements at 
different scales. 
 
\begin{figure} 
\centerline{\psfig{file=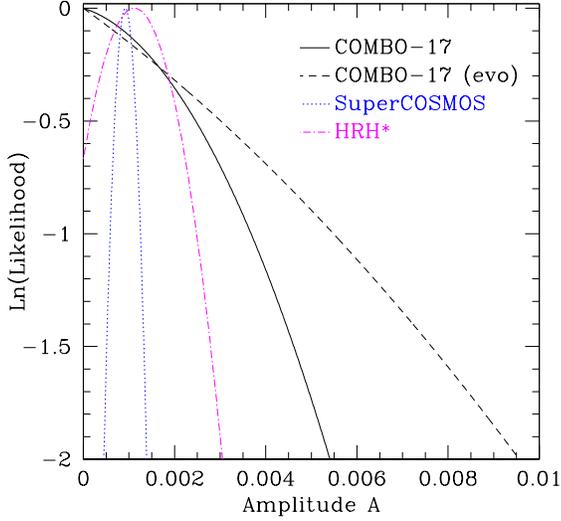,width=7.5cm,angle=0,clip=}} 
\caption{The likelihood of the amplitude A of the model fit
for the intrinsic galaxy alignment correlation function 
equation (\ref{eqn:twopowlaw}), as determined from the COMBO-17  
photometric redshift dataset assuming no evolution galaxy clustering, 
(solid), and assuming stable galaxy clustering, (dashed), 
see Section~\ref{sec:C17evo}. Over-plotted is the likelihood of the amplitude 
from the modified HRH* results, (dot dash), and the likelihood of the
amplitude from the SuperCOSMOS ellipticity variance measurements, (dotted),
as detailed in Section~\ref{sec:newmodel}.}  
\label{fig:likelihood} 
\end{figure}

We calculate the likelihood of our HRH* model from the COMBO-17 
results allowing the amplitude to 
vary, see Fig.~\ref{fig:likelihood}.  We find 
that the best fit is at zero amplitude, and that an HRH* model with an 
amplitude greater than $A = 0.0054$ 
can be rejected with 95\% confidence.  We therefore 
find an observationally constrained maximum amplitude intrinsic alignment 
model such that: 
\be 
\eta_{\rm C17}(r) < \frac{0.0054}{1 + r^2} \qquad {\rm 95\%\ confidence}. 
\label{eqn:max} 
\ee 
Note that using $16\times16$ arcminute
sections in the correlation estimator equation (\ref{corr}), we find
consistent results with a 95\% upper limit $A=0.0039$. 

\subsection{Application of the HH weighting scheme to COMBO-17} 
\label{sec:application} 
 
\begin{figure} 
\centerline{\psfig{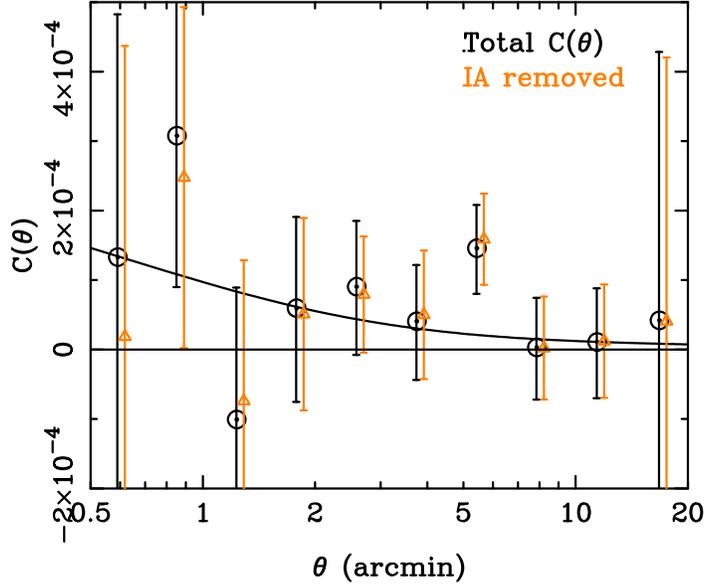}} 
\caption{Application of the intrinsic alignment suppression weighting 
scheme as described in the text. The circles are the total 
correlation function $C(\theta)$ measurements from all galaxy pairs 
whereas the triangles are the $C(\theta)$ measurements 
from just the distant galaxy pairs. The curve plotted is the 
expected weak lensing signal for a $\Lambda$CDM model normalised to 
$\sigma_8=0.8$. } 
\label{fig:weighting} 
\end{figure}

With our constraints on the amplitude of the intrinsic alignment 
signal from the previous section, we can now apply the HH 
weighting scheme  
to the COMBO-17 data. In order to compare with the expected 
$\Lambda$CDM weak lensing 
shear signal, we apply the weighting scheme to the CDFS and S11 fields 
only, in order to avoid the much larger shear signal seen in the A901 
field due to the presence of the 
supercluster \cite{MLB02}. To calculate the optimal 
$\alpha(\theta)$ weights, (see appendix), we 
use the COMBO-17 redshift distribution for the CDFS and S11 fields, 
with our chosen 
input model for the intrinsic alignment signal as given by the upper 
limit of $\eta_{ {\rm C}17}(r)$ equation (\ref{eqn:max}), thereby 
ensuring removal of any feasible intrinsic alignment contamination. 
With values for $\alpha(\theta)$, shown in Fig.~\ref{fig:alpha},  
we calculate the total and distant-pair correlation 
functions as given by equations (\ref{eqn:weightedcor}) and (\ref{corr}), 
the results of which are shown in Fig.~\ref{fig:weighting}. 
The total $C(\theta)$ (circles) will include intrinsic 
alignment contamination, whereas the 
distant-pair signal (triangles) excludes virtually all intrinsic alignment 
at the expense of a small increase in the shot noise. 
Over-plotted is the cosmic shear prediction for a 
$\Lambda$CDM model, equation (\ref{eqn:gg}), 
normalised to $\sigma_8=0.8$, where the  
COMBO-17 photometric redshift distribution for the CDFS and S11 fields
has been used for this calculation. 
 
This result is a good example of the application of the HH weighting 
to a weak lensing data set, effectively removing the systematic 
error from intrinsic galaxy alignments whilst producing only a 
small increase in the shot noise.  Due to the fact that we are 
only using a small subset of the COMBO-17 data set (0.5 square 
degrees limited to magnitude $R<24$), we are unable to use this 
noisy result directly to investigate the effect on cosmological 
parameter estimation.  We can, however, use the likelihood of the 
amplitude of the intrinsic alignment signal to calculate its 
effect on parameter estimation from the full COMBO-17 deep 
$R$-band sample.  We do this in Section~\ref{sec:implications}. 
 
\section{Effect of galaxy clustering evolution} 
\label{sec:evo} 
 
All previous studies of intrinsic galaxy alignment have assumed weak 
evolution in galaxy clustering which becomes important when converting 
measured 3D intrinsic ellipticity correlations, $\eta(r)$, into 
angular correlation functions $C_I(\theta)$, see 
equation (\ref{eqn:Ctheta}).  If there is evolution in galaxy clustering 
such that at high redshift there are proportionally  less nearby 
galaxy pairs than at low redshifts, this will lower predictions of 
$C_I(\theta)$.  The 
clustering evolution of galaxies has often been quantified, although 
without strong theoretical motivation,  
by a redshift-dependent two point correlation function 
of the form \cite{GrothPeeb} 
\be 
\xi_{gg}(r,z) = \left( \frac{r}{r_0} \right)^{\gamma} ( 1 + z) 
^{-(3+\epsilon)}. 
\label{eqn:evo} 
\ee 
Observational studies have found that this simple 
redshift-dependent model provides a poor fit to data 
\cite{Wilson,McCracken}, which is most likely due to the fact that 
different populations of galaxies have different intrinsic clustering. 
Parameters for equation (\ref{eqn:evo}) are therefore 
observationally challenging to determine and as yet fairly uncertain. 
\scite{Wilson} investigates the redshift evolution for a single population of 
galaxies in the UH8K weak lensing fields and 
it is these results that we will use in order to investigate 
the effects of galaxy clustering evolution in intrinsic alignment 
studies such that $\gamma = -1.8$, $r_0 = 5.25 h^{-1}{\rm Mpc}$ and 
$\epsilon=0$.  For $\epsilon=0$, the evolution 
is of stable clustering, where the galaxies are dynamically 
bound and stable at small scales. 
 
\begin{figure} 
\begin{tabular}{c} 
\psfig{file=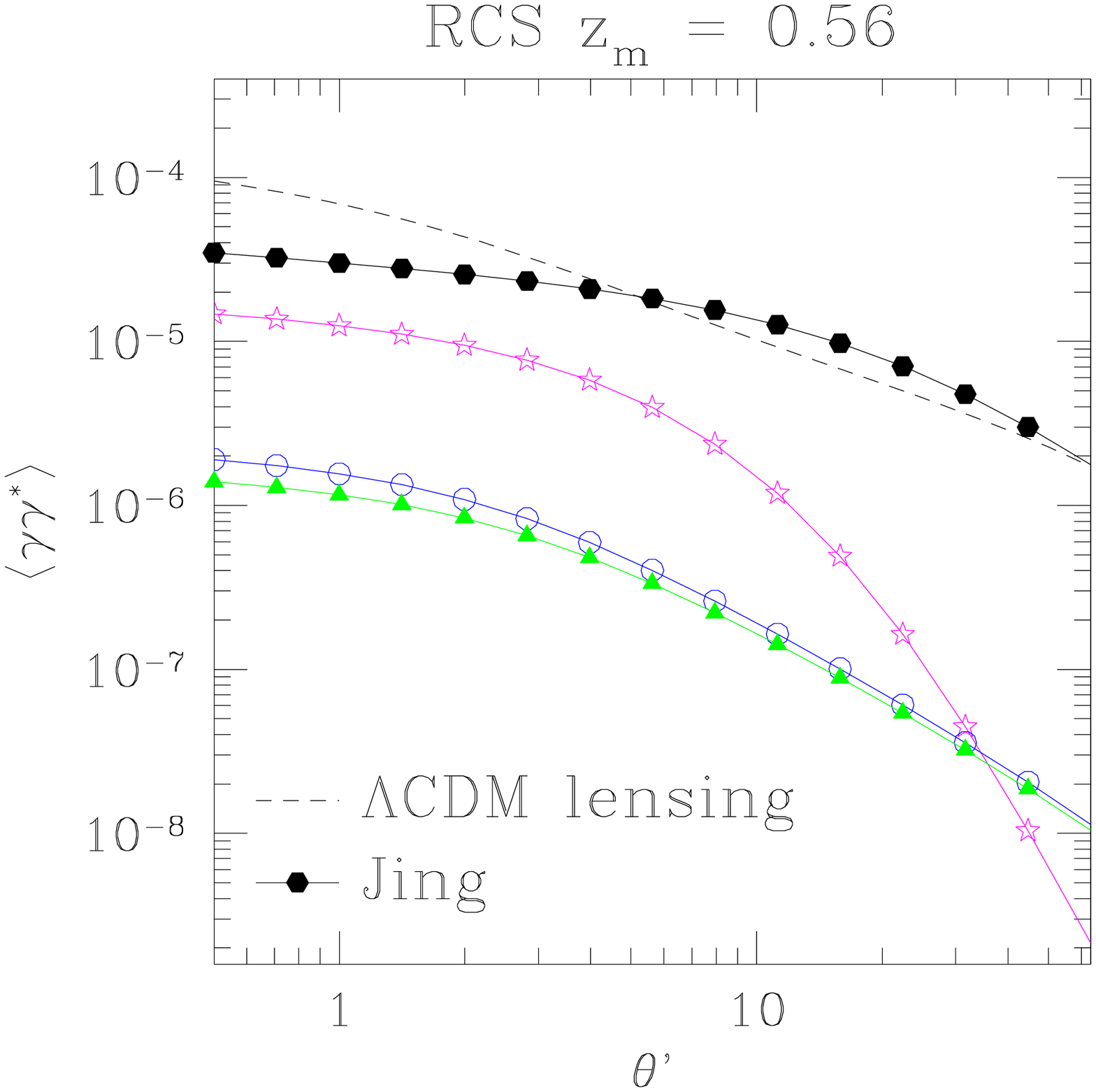,width=7.0cm,angle=0,clip=} \\ 
\psfig{file=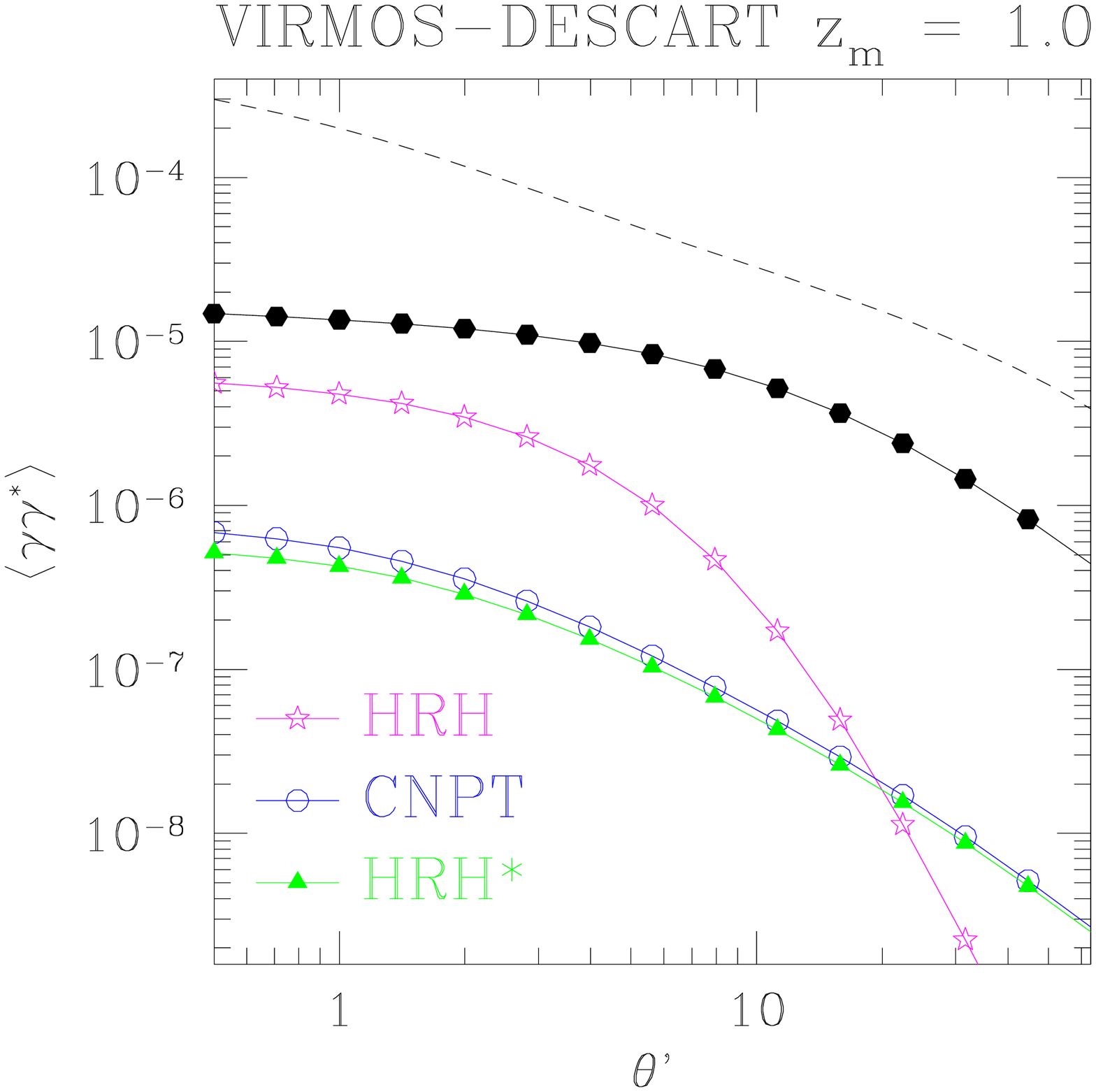,width=7.0cm,angle=0,clip=} 
\end{tabular} 
\caption{Predicted intrinsic alignment angular shear correlation 
functions $\langle \gamma \gamma^* \rangle_I$ including stable galaxy
clustering evolution,  
Jing (filled), HRH (stars), CNPT (circles) and HRH* (triangles). 
The intrinsic alignment contributions are compared to the angular 
shear correlations from $\Lambda$CDM weak gravitational lensing, 
normalised to $\sigma_8 = 0.8$, (dashed), 
for the RCS with depth $z_m = 0.56$, 
(upper) and for the VIRMOS-DESCART survey with depth $z_m = 1.0$, (lower).} 
\label{fig:Ctheta_evo} 
\end{figure}

Fig.~\ref{fig:Ctheta_evo} shows the predicted 
angular shear correlation function, 
$\langle \gamma \gamma^* \rangle_I = \frac{1}{4}C_I(\theta)$ calculated
including galaxy clustering evolution for each 
intrinsic alignment model for the RCS and VIRMOS-DESCART surveys. 
We also plot for comparison the 
shear correlations expected from weak gravitational lensing, 
equation (\ref{eqn:gg}).   Comparing this with Fig.~\ref{fig:Ctheta} 
we note that by assuming stable clustering evolution, the 
intrinsic alignment contribution at small angular scales is 
significantly reduced compared to previous results. 
This is due to the fact that it is 
at small angular scales that high redshift galaxy pairs are 
close enough to contribute to any intrinsic correlations.  If 
galaxies are less strongly clustered at high redshift then there will 
be fewer high-redshift close pairs and hence the angular intrinsic 
correlation function is reduced. 
 
\subsection{Aperture mass B modes} 
\begin{figure} 
\centerline{\psfig{file=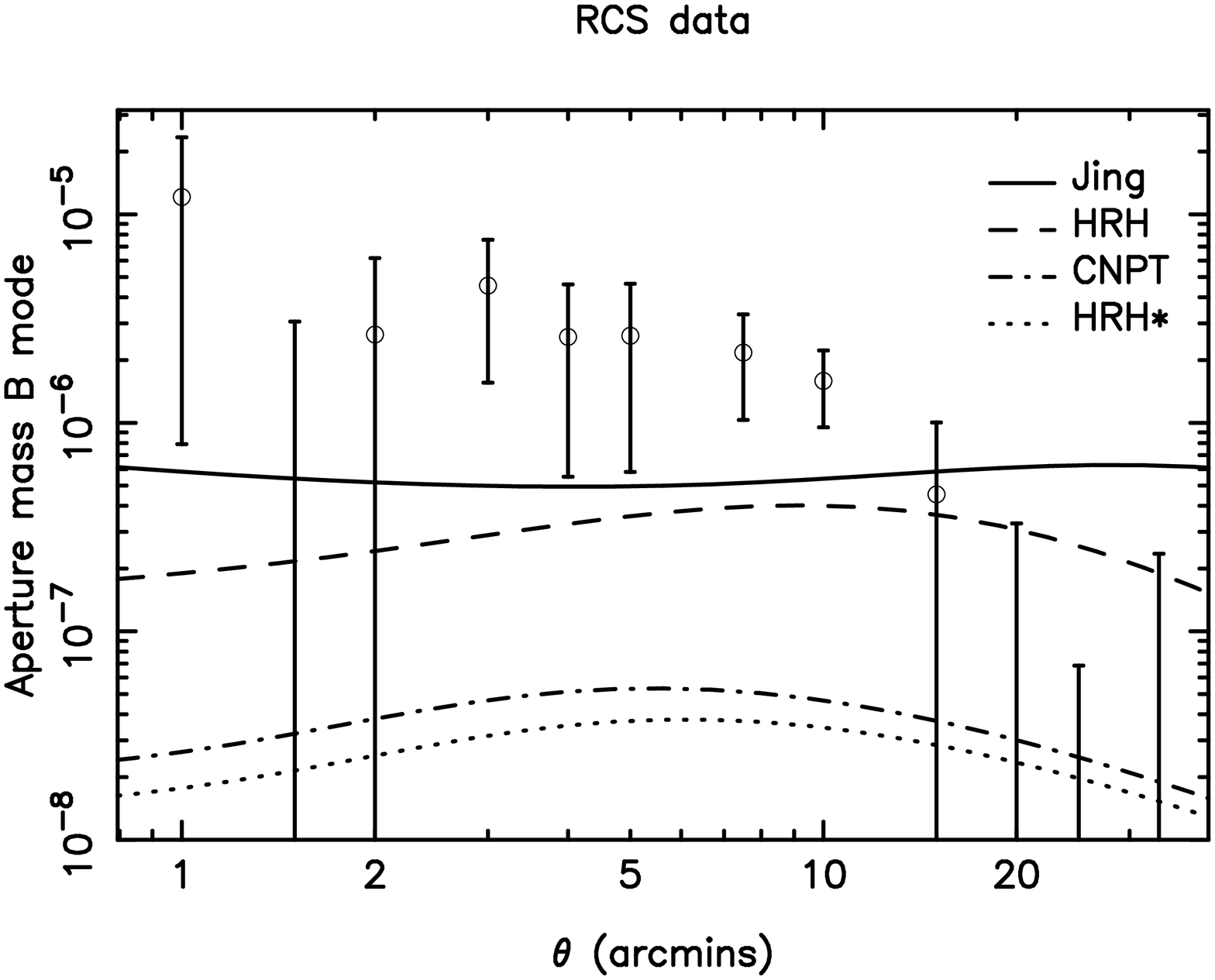,width=7.0cm,angle=270,clip=}} 
\centerline{\psfig{file=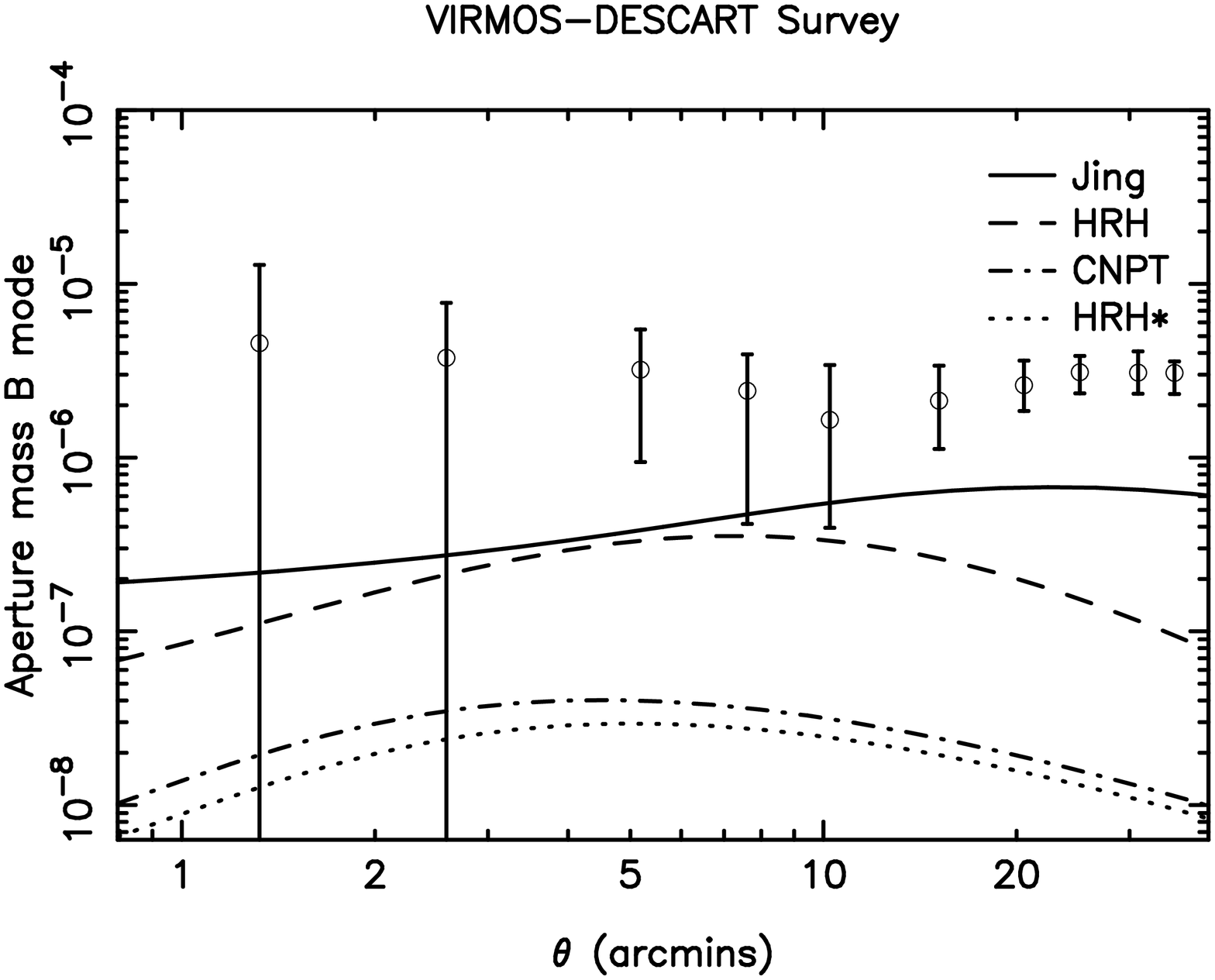,width=7.0cm,angle=270,clip=}} 
\caption{Intrinsic alignment contributions to the aperture mass B mode 
  $\langle M^2_\perp \rangle$ assuming evolution in galaxy clustering; 
  Jing(solid), HRH (dashed) and CNPT(dot dash). The upper panel 
  compares stable clustering intrinsic alignment predictions for the RCS 
  survey with their measured B mode.  The lower panel compares stable 
  clustering intrinsic 
  alignment predictions for the VIRMOS-DESCART with their measured B mode.} 
\label{fig:M_ap_evo} 
\end{figure}

We repeat the analysis of Section~\ref{sec:MapB} with the intrinsic 
alignment correlation functions now calculated including galaxy clustering 
evolution, as shown in Fig.~\ref{fig:Ctheta_evo}.
Fig.~\ref{fig:M_ap_evo} shows these results,  comparing the  
aperture mass B mode measurements reported 
by the RCS, (upper panel), and VIRMOS-DESCART survey, (lower panel), 
and the four different intrinsic 
alignment B modes. 
Unlike the clear result from Section~\ref{sec:MapB} the inclusion of 
stable galaxy clustering reduces the expected B mode contribution from 
intrinsic galaxy alignments, permitting all intrinsic alignment models 
where the observed B modes are considered as upper limits. 
 
\subsection{COMBO-17 Correlation Analysis} 
\label{sec:C17evo} 
\begin{figure} 
\centerline{\psfig{file=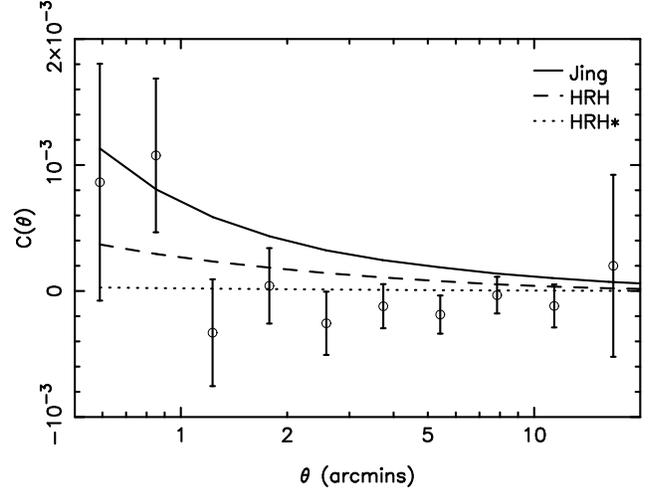,width=7.0cm,angle=270,clip=}} 
\caption{The intrinsic shear correlation function from close galaxy pairs 
  in the COMBO-17 survey, compared to predictions from Jing (solid), 
  HRH (dashed) and HRH* (dotted) for stable clustering galaxy evolution.} 
\label{fig:expairs_evo} 
\end{figure}

We have also investigated the effect of including galaxy evolution in 
the intrinsic alignment models for the COMBO-17 correlation 
analysis. The resulting $C_{\rm IA}(\theta)$  
measurements are shown in Fig.~\ref{fig:expairs_evo} along with the 
predictions for the Jing, HRH and HRH* intrinsic alignment models where 
stable clustering galaxy evolution has been included. Here, we find that, 
although not as clear a result as found in Section \ref{sec:C-17}, we 
can still exclude the Jing model and, once again, the HRH* model is 
favoured.  We calculate the likelihood of our HRH* model which includes 
clustering evolution, allowing the amplitude A to vary, see 
Fig.~\ref{fig:likelihood}. 
We find again that the best fit is at zero amplitude, and that an HRH* 
evolving model with an amplitude greater than $A = 0.0096$ 
can be rejected with 95\% confidence. 
 
\section{Implications for weak lensing measurements} 
\label{sec:implications} 
\subsection{Cosmological parameter constraints} 
 
We now turn to the implications of our intrinsic alignment 
amplitude constraints for 
cosmological parameter estimation.  To do this we follow
\scite{MLB02} using measurements from the deep $R<25.5$ 
COMBO-17 data of the two correlation functions, $C_1(\theta) =
E[\gamma_t\gamma_t:\theta]$ and $C_2(\theta)=
E[\gamma_r\gamma_r:\theta]$ from equation (\ref{corr}), with $W_{ab} =
1$ for all galaxy pairs, in order to obtain a joint measurement 
of the normalisation of the matter power spectrum, $\sigma_8$ and the 
matter density, $\Omega_m$.  We perform a $\chi^2$ fitting 
procedure on the correlation function measurements ordered in a data
vector ${\bf d} = \left\{ C_1(\theta_1),...,
C_1(\theta_n),C_2(\theta_1),..., C_2(\theta_n) \right\}$, 
for a set of theoretical parameters,
$(\sigma_8,\Omega_m,H_0,A)$, where $H_0$ is the Hubble parameter and 
A is the amplitude of the intrinsic
alignment signal in equation (\ref{eqn:twopowlaw}), by calculating
\be
\chi^2 = \left[{\bf d} - {\bf x}(\sigma_8,\Omega_m,H_0,A)\right]^T \, {\bf
V}^{-1}\, \left[{\bf d} - {\bf x}(\sigma_8,\Omega_m,H_0,A)\right], 
\label{eqn:chisqproc}
\ee
where ${\bf x}(\sigma_8,\Omega_m,H_0,A)$ is a theory vector containing the
$C_1(\theta)$ and 
$C_2(\theta)$ correlation functions calculated for the cosmological
model.  ${\bf V} = \langle{\bf d}{\bf d}^T\rangle$ is the
covariance matrix of the data measurements
which can be estimated, assuming that sections of the
same field are uncorrelated, from equation~\ref{covcorr}.
We assume a flat cosmology $\Omega_m + \Omega_\Lambda =
1$ and a nonlinear
CDM matter power spectrum, $P_\delta$, calculated using the fitting
formula from \scite{Rob} with the transfer function of
\scite{HuEisen}, where the initial slope of the power
spectrum is $n=1$. In contrast to \cite{MLB02}, we do not fix
the Hubble constant $H_0$, but marginalise over it, with a Gaussian prior
$p(H_0)$ set by the
WMAP results with $H_0 = 72 \pm 5 \, {\rm km} \, {\rm s}^{-1} \,{\rm
Mpc}^{-1}$  \cite{WMAP}. 

We calculate $\chi^2$ equation (\ref{eqn:chisqproc}), for values of 
$\sigma_8$ ranging from $0.3$ to $1.5$ and values of
$\Omega_m$ ranging from $0.1$ to $1.0$, on a grid with 0.05 spacing in
both these parameters, for two cases
(a) the case where we ignore any effect of intrinsic 
alignments setting $A=0$ and (b) the case where we also vary
values of $A$ from $0.0$ to $0.01$ in steps of $0.0005$ and then
marginalise over the intrinsic alignment amplitude.  For the
marginalisation we use the probability distribution for the 
intrinsic alignment amplitude $p(A)$ obtained from COMBO-17 and shown in
Fig.~\ref{fig:likelihood} such that
\be
p(\sigma_8,\Omega_m) = \int dA \int dH_0\,
p(\sigma_8,\Omega_m,H_0,A)p(H_0)p(A),
\ee
where $p(\sigma_8,\Omega_m,H_0,A) \propto \exp(-\chi^2/2)$.  As in 
Section~\ref{sec:MapB}, we assume that intrinsic galaxy alignments have
no preferred tangential or radial alignment, 
and that $C_{\rm I}(\theta)_{tt} \approx
C_{\rm I}(\theta)_{rr} \approx \frac{1}{2} C_{{\rm I}}(\theta)$, where
$C_{{\rm I}}(\theta)$ is now calculated for the deep $R<25.5$ 
COMBO-17 estimated redshift distribution detailed in \scite{MLB02}.

\begin{figure} 
\centerline{\psfig{file=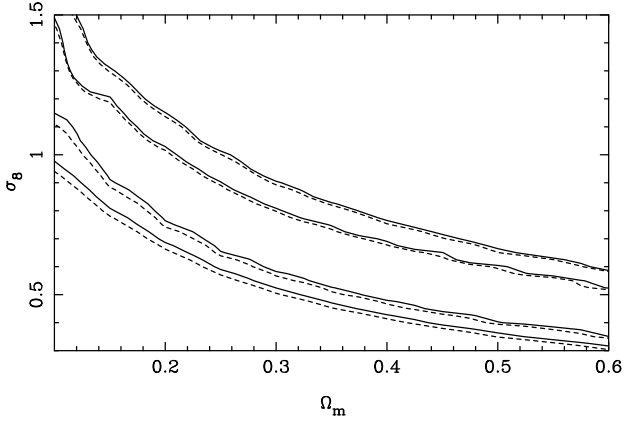,width=5.5cm,angle=270,clip=}} 
\caption{The probability surface for $\sigma_8$ and $\Omega_m$ from 
COMBO-17 as calculated using the shear correlation function 
measurements $C_1(\theta)$ and $C_2(\theta)$ in combination, after
marginalising over the Hubble constant. 
The intrinsic alignment signal is either assumed to be zero (solid 
contours) or is marginalised over (dashed contours). The inner and
outer contours in each case correspond to $\Delta\chi^2=2.3$ and $6.17$. 
confidence regions respectively.} 
\label{fig:sig8om} 
\end{figure}

The resulting constraints in the 
$\sigma_8-\Omega_m$ plane are shown in 
Fig.~\ref{fig:sig8om}. Assuming no 
intrinsic alignment signal, we obtain a best 
fit measurement for the  normalisation of the power spectrum of 
\be 
\sigma_8(\Omega_m/0.27)^{0.6}=0.74 \pm 0.11, 
\ee 
while for the case where we have marginalised over the intrinsic alignment 
amplitude derived assuming no clustering evolution, we find our 
estimate drops to 
\be 
\sigma_8(\Omega_m/0.27)^{0.6}=0.71 \pm 0.11.
\ee 
The errors are larger than quoted in
\scite{MLB02} because we have marginalised over $H_0$ rather than
fixing it at $68$ km$\,$s$^{-1}$Mpc$^{-1}$, but they remain relatively small
in comparison to other surveys, bearing in mind the
COMBO-17 survey area.  This is a result of the 
good determination of source redshifts in the COMBO-17 survey, and
the relatively high number density of resolved background sources used in
this analysis. 
If we marginalise over
the $H_0$ distribution obtained from CMB and 2dF with $H_0 = 68 \pm 5
\, {\rm km} \, {\rm s}^{-1} \,{\rm Mpc}^{-1}$  
\cite{Efstathiou2002,Percival02}, these numbers change to $
\sigma_8(\Omega_m/0.27)^{0.6} = 0.78$ and $0.77$, with an error of
0.12.
Note that all these error bars include a small error added in
quadrature from the imperfect estimation of the covariance matrix
(section 4.1).  
 
\begin{figure} 
\centerline{\psfig{file=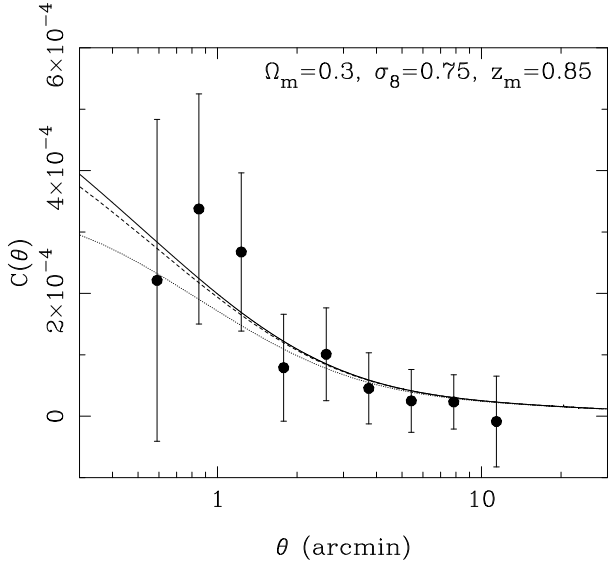,width=8.0cm,angle=0,clip=}} 
\caption{The total shear correlation function for the deep COMBO-17 survey, 
along with the best fit $\Lambda$CDM weak lensing correlation function 
assuming no intrinsic alignments (solid).  Subtracting off the 
the best-fitting intrinsic alignment model (HRH*) decreases the
lensing correlation function to the dashed line, 
and with the $95\%$ upper limit (obtained from the shallower survey 
with photometric redshifts) the lensing correlation function
decreases to the dotted line.} 
\label{fig:cdata} 
\end{figure}

Note that if we instead use measurements of
the total correlation function, $C(\theta)=C_1(\theta)+C_2(\theta)$
we find the same reduction of 0.03 in our estimates of $\sigma_8$. 
With the intrinsic alignment constraints from the SuperCOSMOS survey
we find a reduction in our estimates of $\sigma_8$ of 0.01. 
Fig.~\ref{fig:cdata} shows the COMBO-17 measurement of the shear correlation
function along with the predicted lensing shear correlation 
function with the HRH* intrinsic alignment signal subtracted.

For the case of stable clustering galaxy evolution, the resulting 
angular intrinsic alignment contamination significantly decreases, and 
we find that the constraints in the $\sigma_8-\Omega_m$ plane are 
unchanged after marginalising over the intrinsic alignment signal, 
derived assuming stable clustering galaxy evolution. 
 
\subsection{Implications for future surveys : CFHTLS and SNAP} 
 
The implications for weak lensing surveys is obviously 
dependent, as shown in Section~\ref{sec:evo}, 
on our understanding of galaxy clustering evolution.  Until agreement 
is reached upon the true redshift dependence of galaxy clustering we consider 
the contamination to weak lensing measurements 
derived assuming no evolution in galaxy clustering, which 
should be considered as upper limits as 
redshift evolution will dilute the signal.   For the modified HRH model 
$\eta_{\rm HRH*}(r)$ we find that 
\ba 
\frac{C_I}{ 4 \langle \gamma \gamma ^{*} 
\rangle} ( \theta = 1 \, \rm{arcmin} ) &=& 0.02 \,\, ({\rm CFHTLS}), \nn &=& 
0.01\,\, ({\rm SNAP} ), 
\label{eqn:contamination} 
\ea 
where we have calculated $C_I (\theta)$ 
from equation (\ref{eqn:Ctheta}) with $\phi_z(z)$ as given in 
equation (\ref{eqn:selfunc}), with $z_m = 1.0$ for the CFHTLS wide 
survey, and $z_m = 1.23$ for the SNAP wide survey.  $\langle \gamma 
\gamma ^{*} \rangle$ has been calculated from equation (\ref{eqn:gg}) 
with $\Omega_m = 0.3$, $\Omega_\Lambda = 0.7$, $\sigma_8=0.8$ and 
$\Gamma=0.2$.  The contamination of weak lensing measurements at 
angular scales $\theta > 1$ arcmin is therefore 
less than $2\%$ for CFHTLS, 
and less than $1\%$ for SNAP.  These estimates are
essentially unchanged if we use the SuperCOSMOS measurement of intrinsic
alignments.  Using the 95\% upper limit for the intrinsic
alignment amplitude $\eta_{\rm C17}(r)$, obtained from the COMBO-17 data,
these estimates are multiplied by approximately 5, giving the
95\% upper limits on any possible intrinsic alignment contamination at
10\% for CFHTLS and 5\% for SNAP. 
For comparison, with stable clustering 
evolution, contamination for both surveys is practically negligible, 
less that $0.2 \%$ for CFHTLS, and less that $0.1\%$ for SNAP.   
 
\scite{RefSNAP03} have shown that for SNAP, the application of
redshift tomography as proposed by \scite{Hu99}, increases the
accuracy of cosmological parameter estimates by a factor of about 2.
This method involves splitting the galaxy sample into two or three
redshift slices, and is therefore more susceptible to higher levels of
intrinsic alignment contamination \cite{CM00}.  Using the redshift
slice distributions as proposed by \scite{RefSNAP03}, we have
calculated the intrinsic alignment contamination for the best-fit HRH*
model, and found that for the three SNAP redshift slices with median
redshifts 0.8, 1.3 and 1.9, the contamination at angular scales
$\theta > 1$ arcmin is found to be, 7.0\%, 2.5\% and 0.7\%
respectively.  For the lowest redshift bin, this level of
contamination is significant, especially if we conservatively
consider the 95\% upper limit from our COMBO-17 constraint, 
which increases the
contamination by a factor of approximately 5 to $\sim 35\%$.  It will
therefore be 
important, when using tomography in weak lensing analysis, to apply a
close galaxy pair downweighting scheme in order to remove this
systematic error.  Note that when considering stable galaxy clustering
evolution the contamination is 1\% for the lowest SNAP
redshift bin.
 
\section{Conclusions} 
\label{sec:conc} 
 
The weak correlation of the ellipticities of galaxy images is an 
indicator of gravitational lensing and a powerful tool for the study 
of dark matter on large scales.  In this paper we have used two 
methods to estimate the extent to which this statistic is contaminated 
by the intrinsic physical alignments of galaxies.  Our main conclusion is that 
the effect is relatively small, but not entirely negligible, and, for 
the COMBO-17 survey, leads to a reduction in the derived amplitude of 
mass clustering of around 3\%. 
 
For the brighter part ($R<24$) of the COMBO-17 survey, which has 
photometric redshifts, we removed close pairs from the shear 
correlation analysis, removing the intrinsic alignment signal, as 
described by \scite{HH03}.  Comparing this shear correlation function 
from the distant pairs with that of the close pairs then allows us to 
estimate the intrinsic alignment signal.  We have also placed limits 
on the intrinsic alignment signal from analysis of the aperture mass 
B mode in the RCS and VIRMOS-DESCART surveys.  We find a consistent 
picture that 
this signal is lower than that expected from analysis of numerical 
simulations (\pcite{HRH00,Jing}), but in broad agreement with the 
semi-analytic calculation of \scite{CNPT02}. 
 
We have reanalysed the numerical simulations to include two effects 
which were originally ignored.  These are a misalignment between the 
angular momentum of the baryons and the halo \cite{vdBosch02}, and the 
finite thickness of disk galaxies \cite{CNPT02}.  Both these effects 
reduce the intrinsic galaxy ellipticity correlation function to a 
level similar to that found by \scite{CNPT02}, 
and consistent with the level determined observationally in this
paper and the level measured in the SuperCOSMOS survey \cite{BTHD02}. 
Note that other effects such as gas-dynamical interaction
have not been taken into account, which
could cause the intrinsic alignment signal to be lowered still further.

Having estimated the contribution of intrinsic alignments to the 
brighter part of the COMBO-17 data, we have computed the likelihood 
for the contamination of the whole COMBO-17 sample which extends to 
$R<25.5$.  We compute the probability distribution for $\sigma_8
(\Omega_m/0.27)^{0.6}$, now marginalising over the Hubble constant.
Marginalising over the amplitude of the intrinsic alignment contamination, 
the mass clustering amplitude $\sigma_8 
(\Omega_m/0.27)^{0.6} = 0.71 \pm 0.11$.  Ignoring intrinsic alignments
leads to a systematic overestimate by $0.03$.  
 
From the COMBO-17 results we have also calculated 95\% upper limits
for the expected 
contamination of future surveys, CFHTLS ($<10.0 \%$) and SNAP
($<5\%$).  With our theoretical model, or with the
intrinsic alignment signal estimated from the SuperCOSMOS data, these
predicted limits become 2\% and 1\% respectively.
With current surveys these levels of contamination are
small enough to be neglected,  
but will be significant in the error budget of future 
high-precision weak lensing surveys.  Both CFHTLS and SNAP aim to 
produce accurate photometric redshift estimates for their galaxy sample 
enabling the use of redshift tomography to further improve 
cosmological parameter estimation.  This technique is susceptible to 
significant contamination from intrinsic galaxy alignments due to the 
thin widths of the redshift bins, increasing the proportion of nearby 
galaxy pairs.  We have shown that, even with the 
low amplitude intrinsic alignment model $\eta_{\rm HRH*}(r)$, with the 
proposed redshift distributions for the SNAP 
tomographic redshift bins, the lowest redshift bin will suffer 
contamination $\sim 7\%$, with a 95\% upper limit of $35 \%$ if we 
consider our observationally constrained upper limits for 
$\eta_{\rm C17}(r)$. Since these surveys will have  
photometric redshift information, it will therefore be vital to remove the 
intrinsic alignment signal using the exclusion of nearby galaxy  
pairs as proposed by \scite{HH03} and \scite{KingSch02}.  
 
Our conclusions are affected by the clustering strength of galaxies, 
as this partly determines how many pairs of galaxies which are close 
on the sky are actually physically close together, and susceptible to 
physical interactions which could lead to intrinsic alignments.  The 
results we have presented so far assume that clustering is independent 
of redshift, but for illustration we have also investigated, without 
strong theoretical motivation, an evolutionary model corresponding to 
stable galaxy clustering.  In this case, the effects at high redshift 
are reduced to a negligible level for COMBO-17, CFHTLS and SNAP, but 
could still be important in the case of weak lensing tomography analysis. 
 
In the process of applying weighting schemes as proposed by 
\scite{HH03} and \scite{KingSch02} to future weak lensing surveys, there will 
potentially be some very interesting by-products. 
For example, with large area redshift slices, the method detailed in 
Section~\ref{sec:constrainIA} could be applied in order to determine 
the strengths of intrinsic galaxy alignments as a function of 
redshift, which throughout this paper we have assumed to be constant. In
principle this could be a useful constraint for galaxy formation and 
evolution studies.  With the SuperCOSMOS results and 
our COMBO-17 intrinsic alignment constraint  
favouring an intrinsic alignment model which includes misalignments 
between baryon and halo angular momentum, there is now 
observational evidence indirectly supporting the finding by \scite{vdBosch02}, 
which has important implications for formation of disk galaxies. 
We therefore conclude that although, for studies of weak gravitational 
lensing, the presence of intrinsic galaxy alignments is an 
inconvenience, they are an interesting topic in their own right. 
 
\section{Acknowledgements} 
 
MLB thanks the University of Edinburgh for support during the 
writing of this paper. CW was supported by the PPARC rolling grant in
Observational 
Cosmology at University of Oxford.  The simulations analysed in this paper 
were carried out using data made available by the Virgo Supercomputing 
Consortium (star-www.dur.ac.uk/frazerp/virgo/) using computers based at the 
Computing centre of the Max-Planck Society in Garching at the at 
Edinburgh Parallel Computing Centre.  We are very grateful to Rob 
Smith for providing us with halos from the simulations and his 
non-linear power spectrum fitting formula and code.  We thank Andi 
Burkert, Henk Hoekstra, Paul Allen, Priya Natarajan, Rob Crittenden 
and Stephi Phleps for useful discussions, and the referee
for very helpful comments.

\bibliographystyle{mnras} 
\bibliography{ceh_new} 
 
\appendix 
\section{Optimal galaxy pair weighting}

\begin{figure} 
\centerline{\psfig{file=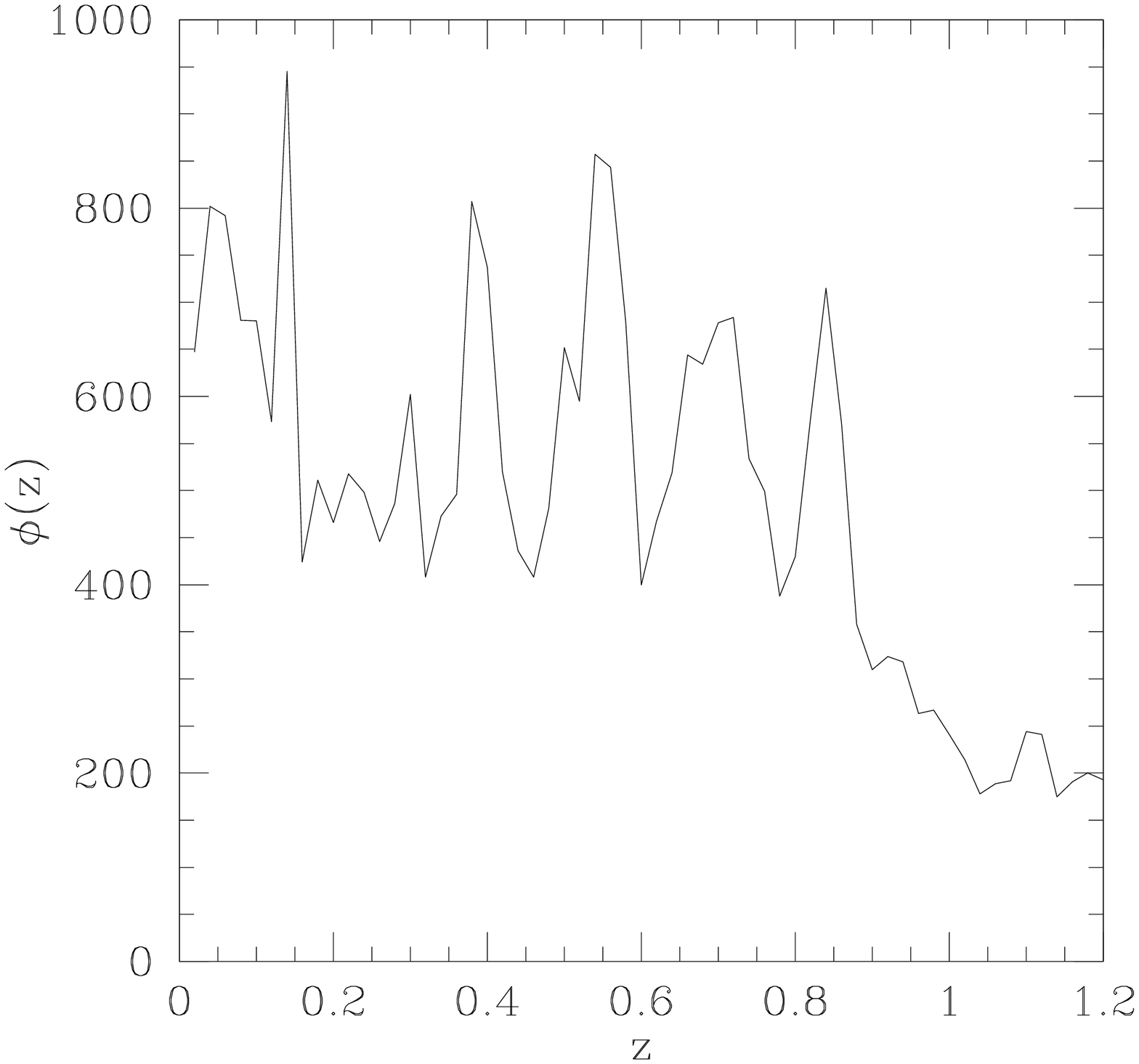,width=7.0cm,angle=0,clip=}} 
\caption{COMBO-17 total redshift distribution for the A901, CDFS 
and S11 fields, limited to $R<24$, with known galaxy cluster members 
removed.} 
\label{fig:C17zs} 
\end{figure}

For a pair of galaxies with estimated redshifts $\hat z_a$ and $\hat 
z_b$ and associated errors $\Delta_z$, following \scite{HH03}, 
we assign a zero weight if $ | \hat z_a - \hat z_b| < \alpha 
\Delta_z $ and a weight of one otherwise.  We choose $\alpha$ to 
minimise the total error on the shear correlation function 
which has two components, one systematic error 
from intrinsic galaxy alignments, $\sigma_{\rm IA}$, and 
one random error from shot noise.  The 
optimum $\alpha$ value will 
depend on angular separation $\theta$, the COMBO-17 redshift 
distribution, $\phi_z(z)$, shown in Fig.~\ref{fig:C17zs}, 
and the median photometric redshift accuracy, 
$\Delta_z = 0.042$.  The shot noise comes from the 
COMBO-17 intrinsic distribution of galaxy ellipticities, $\sigma_e = 
0.67$.  The estimated value for $\sigma_{\rm IA}$ depends on 
which $\eta(r)$ model we choose. 
The conservative approach is to consider the highest 
amplitude intrinsic alignment model, thereby ensuring all possible 
pairs of galaxies that could contribute to the contaminating intrinsic 
alignment signal are removed.  It is this approach that we use in 
Section~\ref{sec:constrainIA} in order to observationally constrain 
the intrinsic alignment contribution to COMBO-17. 
By doing this however there is the expense of a potentially needless high 
residual shot noise, and therefore the preferred approach which we use 
in Section~\ref{sec:application}, is to use the upper limit of 
$\eta_{\rm C17}(r)$ 
in order to optimally remove the intrinsic alignment contamination 
from the weak lensing signal in COMBO-17. 
 
\begin{figure} 
\centerline{\psfig{file=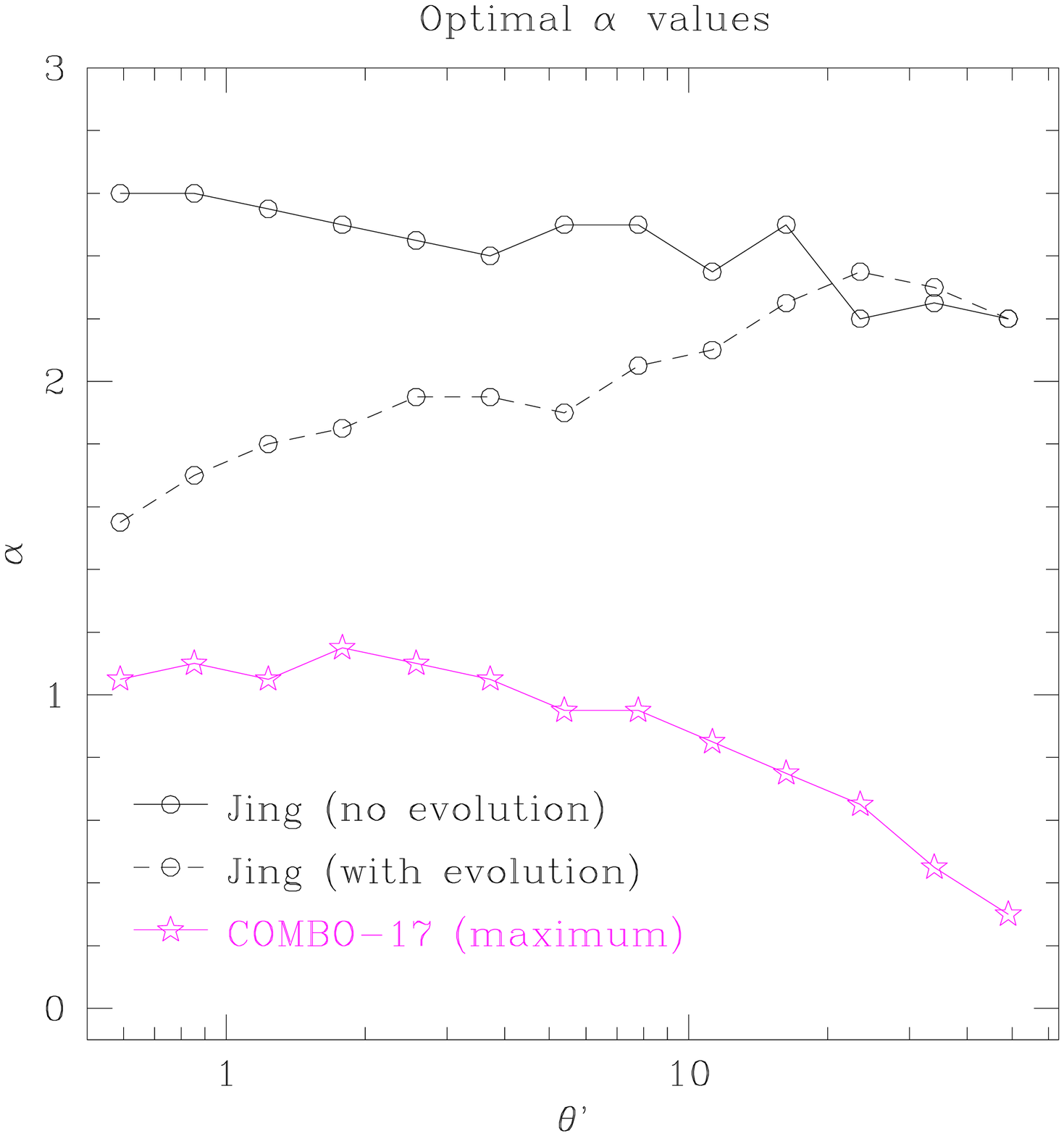,width=7.0cm,angle=0,clip=}} 
\caption{Optimal $\alpha$ values for the COMBO-17 redshift 
distribution for the Jing intrinsic 
alignment model (circles), and the upper limit from the 
COMBO-17 constrained model (stars).   
For the Jing model we show the optimal $\alpha$ 
values considering two different galaxy clustering models: 
zero evolution (solid) and stable clustering (dashed)} 
\label{fig:alpha} 
\end{figure}

Fig.~\ref{fig:alpha} shows the optimal $\alpha$ values calculated 
for the COMBO-17 redshift distribution for the highest amplitude intrinsic 
alignment model: Jing (circles), and the upper limit COMBO-17 model 
(stars).  Spikes in the COMBO-17 redshift 
distribution, shown in Fig.~\ref{fig:C17zs}, mean that 
$\alpha(\theta)$ is not a smooth function. 
The higher amplitude Jing intrinsic 
alignment model requires more galaxy pairs to be rejected than 
for the lower amplitude COMBO-17 model, where 
the intrinsic alignment contribution is of similar amplitude to the 
shot noise.  In this case the best results are obtained with the rejection of 
only the closest galaxy pairs but this would not necessarily be the case 
for large area surveys where high galaxy number counts reduce the 
total shot noise. 
 
We derive $\alpha$ values for the Jing model 
considering no evolution in galaxy clustering (solid) for 
Section~\ref{sec:C17model} and  
considering stable clustering (dashed) for Section~\ref{sec:C17evo}.   
The angular correlation signal from intrinsic galaxy alignments are 
less when we include galaxy clustering evolution, therefore 
the optimal $\alpha$ values are less. 
The expected intrinsic alignment ellipticity 
correlation signal with stable clustering, is 
fairly constant for angular scales $\theta<10$ arcmin 
and this is reflected in 
the optimal $\alpha$ values.   As the shot noise decreases with 
increasing $\theta$, the optimal $\alpha$ can increase to remove more of 
the intrinsic alignment signal without increasing the total error. 
 
\end{document}